\documentclass[11pt]{article}
\usepackage{epsfig}

\newcommand{\R}{${\cal R}^3$}
\newcommand{\absosq}[1]{\bigl\vert#1\bigr\vert^2}
\newcommand{\braket}[2]{\langle#1\vert#2\rangle}
\newcommand{\BOK}[3]{\langle#1\vert#2\vert#3\rangle}
\newcommand{\ket}[1]{\vert#1\rangle}
\newcommand{\sandwich}[3]{\langle#1\vert#2\vert#3\rangle}

\begin{document}
\title{THE ONE, THE MANY, AND THE QUANTUM}
\author{Ulrich Mohrhoff\\
Sri Aurobindo Ashram, Pondicherry 605002, India\\
\tt ujm@auroville.org.in}
\date{}
\maketitle
\begin{abstract}
\noindent The problem of understanding quantum mechanics is in large 
measure the problem of finding appropriate ways of thinking about the spatial 
and temporal aspects of the physical world. The standard, substantival, set-%
theoretic conception of space is inconsistent with quantum mechanics, and so is the 
doctrine of local realism, the principle of local causality, and the mathematical physicist's 
golden calf, determinism. The said problem is made intractable by 
our obtruding onto the physical world a theoretical framework that is more 
detailed than the physical world. This framework portraits space and time as infinitely 
and intrinsically differentiated, whereas the physical world is only finitely 
differentiated spacewise and timewise, namely to the extent that spatiotemporal relations 
and distinctions are warranted by facts. This has the following consequences: (i)~The 
contingent properties of the physical world, including the times at which they are 
possessed, are indefinite and extrinsic. (ii)~We cannot think of reality as being built 
``from the bottom up'', out of locally instantiated 
physical properties. Instead we must conceive of the physical world as being 
built ``from the top down'': By entering into a multitude of spatial relations with 
itself, ``existence itself'' takes on both the aspect of a spatially differentiated 
world and the aspect of a multiplicity of formless relata, the fundamental 
particles. At the root of our interpretational difficulties is the ``cookie cutter 
paradigm'', according to which the world's synchronic multiplicity is founded on 
the introduction of surfaces that carve up space in the manner of
three-dimensional cookie cutters. The neurophysiological underpinnings of this 
insidious notion are discussed.
\end{abstract}
\pagebreak

{\leftskip=1.2in\noindent
``I feel that the real joke that the eternal inventor of enigmas has presented us 
with has absolutely not been understood as yet.''\par}
\begin{flushright}
Albert Einstein~\cite[p. 411]{Pais1982}
\end{flushright}

\section{Introduction}

Quantum mechanics is an incredibly successful theory. No experiment, no 
observation, has ever given the lie to it. But if it is not lying, what does it tell us 
about the world? A fairly typical answer is: Nothing; quantum mechanics 
concerns statistical regularities in the behavior of measuring instruments; any 
attempt to go beyond the ``brute facts'', to give a realistic or epistemological 
account of how it is that the statistical regularities predicted by quantum 
mechanics come out the way they do, is idle metaphysics. Considering the 
widely divergent and more or less bizarre accounts that abound, this resigned 
attitude appears justified. But is it?

In this essay I examine the metaphysical presuppositions that stand in the way 
of making sense of quantum mechanics and trace them to their 
neurophysiological underpinnings. Because these underpinnings to a 
considerable extent determine the nature of the phenomenal world and 
consequently prejudice our thinking about the physical world, rejecting those 
presuppositions is no easy task. But it is worth the effort, for at the end we 
shall find that the mathematical elegance and simplicity of quantum mechanics 
is matched by the depth and transparency of its ontological message.

The problem of understanding quantum mechanics is in large measure the 
problem of finding appropriate ways of thinking about the spatial and temporal 
aspects of the physical world. Section~2 contrasts the salient features of 
phenomenal space with the standard mathematical description of physical 
space. It is argued that the standard description has 
empirically unwarranted features that stand in the way of finding the right 
interpretation. An alternative, essentially relationalist way of thinking 
is proposed.

In the sections that follow the relevant empirical findings are progressively 
taken into account. Section~3 takes account of the existence of noncomposite 
entities, a.k.a. the fundamental particles of matter. Its chief conclusion is that 
these entities must be thought of as formless. The still prevalent notion that a 
fundamental particles has a (pointlike) form is an unwarranted (in fact, 
illegitimate) importation from the phenomenal world into the world of physics. 
The form of a material object is the set of its internal spatial relations. An object 
that lacks internal spatial relations also lacks a form.

Section~4 takes account of the behavior of indistinguishable particles. It is 
argued that a fundamental particle is not something to which existence is 
contingently attributable. The basic reality of matter is a single entity, and this 
entity cannot be qualified as anything more particular or determinate than 
``existence itself''. Considered in relation to each other, fundamental particles 
are instances of ``existence itself''. Considered out of relation to other particles, 
each fundamental particle is ``existence itself''. The individuation or multiple 
instantiation of this entity is discussed.

Section~5 takes account of the behavior of electrons in two-slit interference 
experiments. This behavior is Nature's corroboration of the relationalist 
conception of physical space developed in Secs.~2 and~3. It is argued that the 
standard, substantival, set-theoretic conception of space is as inconsistent with 
quantum mechanics as absolute simultaneity is with special relativity. So, 
therefore, is the doctrine of local realism, the principle of local causality, and 
the mathematical physicist's golden calf, determinism.

Section~6 deals with two frequent mistakes, the error of the instrumentalist, 
who rejects all efforts to make ontological sense of quantum mechanics, and 
the error of the quantum realist, who considers probability one sufficient for the 
existence of an element of reality. While the instrumentalist errs by ignoring the 
possibility of a quantum world with extrinsic properties, the quantum realist errs 
by ignoring that the properties of the quantum world, including the times at which 
they are possessed, are extrinsic: They ``dangle'' from what happens or is the 
case in the rest of the world. The contingent properties of the quantum world 
exist precisely to the extent that their values are indicated.

Section~7 pinpoints the sense in which the relative positions of the world's 
material constituents are indefinite. This involves reference to objective 
probabilities, which can be assigned only to counterfactuals. The very 
possibility of assigning objective probabilities to the possible results of an 
unperformed measurement entails that the possessed values of 
quantum-mechanical observables are 
extrinsic. The extrinsic nature of the world's contingent properties thus follows 
directly from their indefiniteness. Hence unless we are willing 
to take seriously the extrinsic nature of quantum-mechanical properties, we 
shall not be in a position to make proper sense of the indefiniteness that is the 
hallmark of quantum mechanics, and hence of quantum mechanics itself.

While even a macroscopic object (defined in Sec.~7) has a position only 
because of the facts from which this can be inferred, the dependence of 
its position on position-indicating facts is a qualitative (ontological or existential) 
dependence, not a quantitative one. For the quantitative purposes of physics, it 
is legitimate to ignore this dependence, to consider macroscopic positions as 
intrinsic, and to apply to them classical causal concepts. This is fortunate, for 
otherwise quantum mechanics would be inconsistent, inasmuch as its very 
formulation presupposes a classical domain of positions that can be thought of 
as being factual {\it per se}. Causality nevertheless is emergent; it is not part of 
the ontological foundation. All the determining that goes on in the physical 
world is the determining of probabilities associated with possibilities. There 
aren't any causally determined facts. The quantum formalism concerns 
probabilistic correlations among facts -- diachronic correlations between facts 
indicating properties of the same system at different times and synchronic 
correlations between facts indicating properties of different systems in 
spacelike separation,~-- and it is these correlations that explain why causal 
explanations work to the extent they do. Trying to causally explain the 
correlations would be putting the cart in front of the horse. All of this is 
explained in Sec.~8.

The seemingly intractable problem of understanding quantum mechanics is 
largely due to our obtruding onto the physical world a theoretical framework 
that is more detailed than the physical world, in that it portraits space and time 
as infinitely and intrinsically differentiated. Section~9 shows that the physical 
world is only finitely differentiated spacewise and timewise, namely to the 
extent that spatiotemporal relations and distinctions are warranted by facts. As 
a consequence, we cannot think of reality as being built ``from the bottom up'', 
on an infinitely and intrinsically differentiated space, out of locally instantiated 
physical properties. Instead we must conceive of the physical world as being 
built ``from the top down'': By entering into a multitude of spatial relations with 
itself, ``existence itself'' takes on both the aspect of a spatially differentiated 
world and the aspect of a multiplicity of relata -- the fundamental particles. By 
allowing the spatial relations to change, it takes on the further aspect of a 
temporally differentiated world.

The final section homes in on the idea that is at the root of our interpretational 
difficulties, traces its philosophical fallout, and examines its neurophysiological 
underpinnings. This is the ``cookie cutter paradigm'', according to which the 
world's synchronic multiplicity is founded on the introduction of surfaces that 
carve up space in the manner of three-dimensional cookie cutters.

\section{The Reality of Spatial Continuity}

The problem of understanding quantum mechanics is in large measure the 
problem of finding appropriate ways of thinking about the spatial and temporal 
aspects of the physical world. The way we think about space is obviously 
indebted to our awareness of phenomenal space, the three-dimensional 
expanse which contains both sensory percepts (a.k.a. qualia or introspectible 
properties) and visual images.\footnote{%
Neuroscience is increasingly backing the notion that experience and 
imagination share the same space. There is mounting evidence that visual 
perception and visual imagery compete for the same processing mechanisms, 
and that the neural processes which produce visual percepts and those which 
produce visual images are to some extent the same~\cite{Finke1980,
Finke-Shepard1986,Shepard-Cooper1982}.}
One of the things that quantum mechanics is trying to tell us is that the way we 
are accustomed to think about space is more appropriate for dealing with the 
spatial aspect of the phenomenal world than it is for dealing with the spatial 
aspect of the physical world. In the following sections I will successively 
introduce the relevant empirical data and examine their implications. In the 
present section I focus on what we can learn by paying attention to our direct 
awareness of phenomenal space and by contrasting it with the standard 
mathematical representation of physical space.

Physicists routinely represent space as a transfinite set of point individuals in 
one-to-one correspondence with triplets of real numbers. This mathematical 
representation lacks certain features that are vital for interpreting quantum 
mechanics correctly, and it possesses certain empirically unwarranted features 
that stand in the way of finding the correct interpretation. The features that are 
lacking are the {\it quality} of continuous spatial extension and a {\it unity} that 
goes beyond the unity of a set, defined by Cantor~\cite[p.~204]{Cantor1932} as 
``a Many that allows itself to be thought of as a One''. The empirically 
unwarranted features are the {\it multiplicity} that is inherent in the set \R\ of all 
triplets of real numbers and the {\it intrinsic distinctness} of the members of \R.

Consider the visual image of a line. This is in an obvious sense continuously 
extended in one spatial dimension. If the line is unbounded, we can set up a 
one-to-one correspondence between the set $\cal R$ of real numbers and 
points on the line; if it is bounded, we can set up a one-to-one correspondence 
between any finite interval of $\cal R$ and points on the line. What is important 
is that these points do not exist in advance of any one-to-one correspondence 
that we may set up. They are introduced into the line by our setting up such a 
correspondence. The line is in an obvious sense divisible into segments, but 
nothing in its image warrants the notion that it is intrinsically multiple, let alone 
that it is a concatenation of point individuals with the cardinality of the real 
numbers. While between two distinct real numbers $a$, $b$ there exists a 
nondenumerable set $S$ of real numbers, between the points that correspond 
to $a$ and $b$ there exists a perfectly continuous and intrinsically 
undifferentiated line segment $L$. $S$ and $L$ are entirely different things. 
$S$ possesses something that $L$ lacks, namely the multiplicity of a 
nondenumerable set and the distinctness of its members. $L$ possesses 
something that $S$ lacks, namely the quality of continuous spatial extension 
and the unity of an intrinsically undivided whole.

Unfortunately, we are so accustomed to conflating the continuity of phenomenal 
space with the discreteness of \R\ that it has become almost impossible for us 
to tease them apart. We tend to visualize the reals as a continuous line without 
realizing that the very act of visualization introduces a qualitative element that 
is not warranted by the mathematical construction of the reals. Conversely, in 
our attempt to get a conceptual grip on physical space we seize on the reals as 
a set that appears to contain sufficiently many elements to ``fill'' a continuous 
line and thus to possess its continuity. We even deprive ourselves of the words 
that are needed to distinguish between $S$ and $L$, as when we use 
the adjectives ``continuous'' and ``discrete'' to qualify {\it sets}. In what follows, 
all words starting with the letters \mbox{``c-o-n-t-i-n-u''} will signify (possession 
of) that continuity which is an obvious pre-theoretical feature of our visual 
percepts and images, and ``discrete'' will refer to that discreteness which is an 
obvious property of every ``Many that allows itself to be thought of as a One''.

Continuity, so defined, is a feature of objects in phenomenal space (that is, of 
visual percepts and visual images), and it is not a feature of any mathematical 
set. Is continuity a feature of objects in physical space?\footnote{%
The phraseology which depicts space as some kind of container in which 
objects are situated is common enough, but it must not be taken literally. 
Phenomenal space contains neither in the set-theoretic sense of 
``containment'' nor in the sense in which a closed surface $b$ encompasses its 
interior. Neither of these senses is applicable to an unbounded continuous 
space. Hence if we say of something that it exists or is situated ``in'' space, 
what we mean is that it is spatially extended, or that it is spatially related to 
something else, or that it stands out from a spatial expanse like the visual 
image of a point.}
Does it exist in the physical world? These questions are quite analogous to the 
following: Does the turquoise of a 
tropical lagoon exist in physical space (that is, is it an objective property of the 
lagoon)? Few would venture an affirmative answer. Qualia belong to 
phenomenal space; they are correlated with certain physical quantities but are 
themselves not physical.\footnote{%
An accurate physical correlate of color sensations is given by the triplets of 
measurable quantities known as ``integrated reflectances''~\cite{Land1977}.}
But the continuity of phenomenal space is as much a {\it qualitative} feature of 
our visual percepts as is the sensation of turquoise. It eludes mathematical 
description just as definitely as does the color of the lagoon. One theorist who 
was acutely aware of this was Hermann Weyl~\cite[p.~23]{Weyl1970}:
\begin{quote}
\ldots\ it ought to be 
emphasized just how little mathematics can claim to capture the phenomenal 
[{\it anschauliche}] nature of space: nothing in geometry concerns what makes 
phenomenal space what it distinctively {\it is}\ldots. Our conceptual theories are 
capable of penetrating only one aspect of space, and only the most superficial 
and formal at that.
\end{quote}
A triplet of real numbers is not the 
same as a point. The difference between two real numbers is not the same as 
the distance between two points. Distances possess, in addition to their values, 
a purely qualitative aspect, and nobody lacking our pre-mathematical grasp of 
phenomenal space is in a position to know it, anymore than Mary, confined 
from birth to a black-and-white room, was in a position to know 
color~\cite{Jackson1986}.

It might be concluded that we should think of continuity in the same way as we 
think of color sensations -- as an appearance or a secondary quality, rather 
than as an actual feature of the physical world: Continuity is a property of 
percepts and images in phenomenal space; the real thing, physical space, is 
discrete and cardinally equal to \R. However, in view of the interpretational 
difficulties associated with quantum mechanics, it may be worth a try to consider 
the continuity we find in phenomenal space, and the undifferentiated unity that 
goes with it, as objective features of the physical world, and to determine how, 
this being so, the multiplicity of \R\ and the distinctness of its members relate 
to the physical world.

It is worth noting, to begin with, that if we conceive of physical space as a set of 
points corresponding one-to-one to the members of \R, the individual points of 
space are not visualizable. If we imagine a point, we imagine something that 
stands out from an otherwise undifferentiated spatial expanse. If we think of 
this point as one of the points of space, then this particular point of space 
possesses a quality that is not present at those points of space that make up 
the surrounding expanse. Hence what we imagine is not a point of space but 
this quality situated or instantiated at a particular point of space. In other 
words, what we imagine is a pointlike {\it object}. We can visualize pointlike 
objects, but we cannot visualize an individual point of space. The points of 
space are not objects but nonvisualizable {\it positions} at which visualizable 
objects may be situated or visualizable qualities may be instantiated.

Positions in phenomenal space can only be defined relatively, as spatial 
relations that hold among visual percepts or images. We have neither empirical 
nor theoretical reasons to doubt that the same is true of physical space: 
Positions are relatively defined, as spatial relations that hold either among 
material objects or among the points of space. The latter alternative, however, 
introduces into the physical world a spatial multiplicity and a degree of spatial 
differentiation that not only are empirically unwarranted but also violate the 
Identity of Indiscernibles, a principle of analytic ontology which says that two 
things cannot have exactly the same properties. Since no objective 
differences correspond to the differences between the triplets of real 
numbers by which we label the points of space, these points have identical 
properties and therefore cannot be thought of as distinct individuals.

The points of space being positions at which material objects may be situated 
or physical qualities may be instantiated, and positions being relatively defined, 
the points of space are relatively defined. We do not have, on the one hand, a 
set of preexistent positions (cardinally equal to the reals) and, on the other 
hand, a set of spatial relations that hold among these positions. The positions 
{\it are} the spatial relations, or else they are defined by them: The exist, as 
distinct locations, by virtue of the distinguishing relations. 
The relata owe their existence to the relations, and the relations owe their 
existence to the material objects or instantiated physical properties to which 
they are attributable.

It follows that the synchronic multiplicity of the world does 
not go beyond the multiplicity of material objects (or locally instantiated 
physical properties\footnote{%
The qualifier ``locally instantiated'' does not imply that locations exist in 
advance of instantiation.}%
) existing at any one time, and the multiplicity of their relations. It does not 
include the nondenumerable multiplicity of the ``points of space''. The 
function of this multiplicity -- the multiplicity of coordinates -- is not to label 
``vacant'' positions but to quantify the undifferentiated spatial relations that 
exist between material objects. (The actually existing spatial relations, which 
spatially differentiate the world, are not themselves spatially 
differentiated.)

This is relationism, the doctrine that space is a family of spatial relations 
holding among the world's material constituents, rather than an additional 
constituent of the world. The alternative to relationism is substantivalism. 
According to the latter, there is such a thing as ``space itself'', and this has, by 
itself, a definite number of dimensions and a metric. The dimensionality of 
physical space, however, is fully determined by the system of spatial relations 
that hold among material objects; it is the number of coordinates needed to 
specify those relations. There is no need to attribute it to a separate constituent 
of the world.

What about the metric? This too need not be attributed to a separate 
constituent of the world. As was pointed out by Riemann~\cite[p.~101]{Weyl1970} 
\cite[p.~752]{Grunbaum1973}, unless physical 
space is inherently discrete (and therefore a separate constituent of the world 
with inherent properties), metric relations are extrinsic to it. On the relationist 
view of space, metric relations exist only between material objects, and the 
assignment of values to such relations is based on the behavior of material 
objects, rather than on properties intrinsic to space. Since the behavior of 
material objects finds formal expression in laws that make reference to 
distances, it is possible that the metric properties of the world and the laws of 
physics are individually underdetermined. What is 
observable, or factual, is the behavior of material objects. To describe it, we 
use both a metric and a set of dynamical laws, neither of which is separately 
observable. Hence, in principle, the same behavior is describable in relation to 
different metrization schemes, by alternative sets of physical laws. This means 
that the metric properties of the world are partly based on a conventional 
choice between alternative sets of laws, as was stressed by 
Poincar\'e~\cite{Poincare1952}.\footnote{%
Arguably, the specific lawful behavior responsible for the metric properties of 
the world consists in the invariant periodicities that are associated with the 
inertial masses of particles~\cite{Anandan1980,Silva1997}.}

If we embrace relationism, we cannot attribute continuity to physical space as if 
this were a separate constituent of the world -- a three-dimensional expanse 
existing independently of its material ``content''. If there is anything in the 
physical world to which we can attribute the continuity that we find in phenomenal space, it 
is the spatial relations that exist between material objects. (As the following 
section will show, it is not necessary to attribute continuity separately to the forms of 
spatially extended objects, for such forms are simply sets of spatial relations.) 
Moreover, since the synchronic multiplicity of the world is limited to the multiplicity of 
material objects existing at any one time and the multiplicity of their relations, each 
relation possesses not only this continuity but also the unity of an undifferentiated 
line segment.

Let me amplify. Given Cartesian coordinates, the spatial relation between two 
material objects $A$ and $B$ is essentially the distance ${\cal D}(AB)$ 
between $A$ and $B$. ${\cal D}(AB)$ is {\it quantitatively} determined by the 
algebraic differences between the coordinates of $A$ and $B$. But the 
distances between material objects also have a {\it qualitative} character. They 
are not just numbers; they are {\it spatial}. We tend to think of ${\cal D}(AB)$ 
as a quantity, and we tend to attribute to this ``quantity'' the multiplicity of 
some interval of the so-called ``real line''. That is, we tend to take it for granted 
that there are as many places between the two objects as there are real 
numbers between $0$ and $d(AB)$, the value of ${\cal D}(AB)$. In reality there 
are as many places in the material world as there are material objects. The 
places at which objects may be located do not actually exist unless objects are 
actually located there. Thus if we consider the distance ${\cal D}(AB)$, we are 
looking at something that possesses the quality of spatial extension as well as 
a value $d(AB)$, but that is destitute of multiplicity. Unless there are other, 
appropriately situated objects, there are no places between $A$ and $B$.

It thus is a mistake to think of ${\cal D}(AB)$ as intrinsically multiple. 
There is no multiplicity that would qualify as inherent in a single 
spatial relation. There are no points or places between $A$ and $B$ unless 
other material objects are situated between $A$ and $B$. It takes a third object 
$C$ to introduce another two distances ${\cal D}(AC)$ and ${\cal D}(CB)$ such 
that $d(AB)=d(AC)+d(CB)$. The same equation does not hold among the three 
distances ${\cal D}(AB)$, ${\cal D}(AC)$ and ${\cal D}(CB)$. None of these 
distances is the sum of anything. ${\cal D}(AB)$ is not a quantity that is ``made 
up'' of quantities; it is a relation that possesses (i)~the qualitative property of 
undifferentiated continuous extension and (ii)~the value $d(AB)$. It interposes 
no places between $A$ and $B$. If anything interposes a location between the 
location of $A$ and the location of $B$, it is another material object $C$.

To thinkers from Aristotle to Kant and Gauss it appeared self-evident that 
points on a line are extrinsic to the line, in the sense that they are additional 
features not contained in its image. They considered the line itself and, by 
implication, space itself is an inherently undivided ``whole'' existing in an 
anterior relationship to limits and divisions. ``Space is essentially one'', Kant 
wrote~\cite[p.~25]{Kant1781}, ``the manifold in it\ldots\ arises entirely from the 
introduction of limits.'' Kant was right in saying that what is continuous is intrinsically 
one. But he was wrong in attributing the world's synchronic multiplicity to the 
introduction of limits. Synchronic multiplicity owes its reality to the spatial 
relations that exist between objects. Therefore there exists no such tension of 
contrast as that between the unity of a three-dimensional continuous expanse 
and a multiplicity of divisions that somehow appear in it. Continuity and unity 
are not properties of ``space itself'' but pertain to (are instantiated with) each 
spatial relation between a pair of material objects. The continuous unity of 
${\cal D}(AB)$ thus in no way conflicts with the spatial multiplicity of the world, 
for the former belongs to {\it each} spatial relation while the latter belongs to 
the totality of spatial relations that exist in the world.

We have assumed the existence of Cartesian coordinates. Since the central 
purpose of this article is to find a way of thinking about the world's actual multiplicity 
that is consistent with quantum mechanics (or a way of thinking about 
quantum mechanics that is consistent with the world's actual multiplicity), this 
assumption is justified by the fact that quantum mechanics {\it presupposes} 
the use of Cartesian coordinates. Specifically, every quantization scheme that 
leads to Hilbert space vectors and Weyl operators presupposes the ``flat'' 
metric that goes with Cartesian coordinates~\cite{Klauder1997, Klauder1998, 
Klauder1999}.\footnote{%
``This assumption [of replacing classical canonical coordinates by 
corresponding operators] is found in practice to be successful only when 
applied to the dynamical coordinates and momenta referring to a Cartesian 
system of axes and not to more general curvilinear coordinates'' -- 
Dirac~\cite[p.~114]{Dirac1947}.}
To see why this should be so, we must bear in mind that in quantum mechanics 
coordinates do not represent positions but {\it values} that are available for 
attribution to the positions of material objects. Since these positions are relative 
positions, a useful coordinate system has to be riveted to some material object 
(e.g., the nucleus of an atom, the center of mass of a composite object, a 
macroscopic part of a macroscopic apparatus). A coordinate system $C_A$ 
riveted to a given object $A$ represents the ``space'' of values that are 
potentially attributable to the positions of objects $B_i$ relative to $A$, but it 
does {\it not} represent the ``space'' of values that are potentially attributable to 
the positions of the objects $B_i$ relative to each other. The appropriate value 
``space'' for the positions of the objects $B_i$, $i\geq0$, relative to $B=B_0$ is 
not $C_A$ but $C_B$.

This would be of no consequence if the relative positions of material objects 
had definite values, for then $C_B$ could be obtained by simply translating 
(and maybe rotating) $C_A$. But since the relative position of each pair of 
material objects is to some extent ``fuzzy'' -- a notion that will be substantiated 
and made precise in what follows,~-- there isn't any point transformation that 
takes us from $C_A$ to $C_B$. It follows that the coordinate ``space'' 
presupposed by quantum mechanics can have nothing to do with curvature. 
This ``space'' of attributable values is always riveted to a {\it single} material 
object $A$, whereas a test for curvature involves at least two different, not 
determinately related, value ``spaces''. (To detect curvature near $A$, one 
needs to know not only the positions of nearby objects $B_i$ relative to $A$ 
but also the positions of the objects $B_i$ relative to each other.) Each individual value 
``space'' is therefore necessarily and trivially flat.

In this section arguments supporting the following notions have been presented:

\begin{itemize}
\item Physical space is not a set of point individuals but a system of spatial 
relations that hold among material objects.

\item Like the sensation of turquoise, continuous extension is a qualitative 
feature of our visual percepts. Unlike that sensation, it is an objective feature of 
the physical world (that is, it is an objective feature of every spatial relation).

\item The relative position of any pair of material objects (or the distance 
between the two objects) lacks multiplicity; it possesses the quality of 
continuous spatial extension (in one dimension) and the undifferentiated unity 
of an unsegmented line in phenomenal space.

\item The spatial multiplicity of the world is the multiplicity of spatial 
relations that hold among material objects. It is not due to the introduction of 
boundaries.

\item A clear distinction has to be made between physical space and the set 
\R\ of values that are potentially attributable to the relative positions of material 
objects. The quality of continuous spatial extension is not attributable to this 
set, nor is the concept of curvature applicable to it.
\end{itemize}

As will become clear in what follows, the conceptual framework staked out so far is 
eminently suited to the task at hand -- making sense of quantum mechanics.

\section{Are Fundamental Particles Pointlike?
\\The Ontology of Synchronic Multiplicity}

Let us now take account of the relevant empirical findings. We begin with the 
existence of simple or noncomposite entities, which we will refer to as 
``fundamental particles''. According to the current standard model of 
elementary particle physics, the particles that are fundamental are the quarks 
and the leptons, but all we need to know or assume at this point is that 
noncomposite objects exist. Combined with the conclusions reached 
in the previous section, this warrants the following claims:

\begin{itemize}
\item Any composite material object (with the possible exception of the universe 
as a whole) is made up of a finite number of fundamental particles {\it and} the 
spatial relations that hold among them. To the particles it owes its finite 
multiplicity; to the spatial relations it owes its spatial extension.

\item There are not two kinds of spatially extended things, such as regions of 
space and a material stuff that occupies or fills them. The only kind of spatial 
extension in existence is the spatial extension that is attributable to possessed 
spatial relations. There is no material stuff that occupies or fills space. What is 
spatially extended either is a spatial relation or has spatial relations among its 
constituents. A composite object ``occupies'' space only in the sense that it is 
partly constituted by the spatial relations between its material constituents.

\item There are not two kinds of synchronic multiplicity, such as a multiplicity of 
material things and a multiplicity of self-existing locations. There is only the 
multiplicity of fundamental particles and the multiplicity of their spatial 
relations. The actually existing locations are the (relatively defined) positions 
that are attributable to fundamental particles and composites thereof.
\end{itemize}

In respect of any given object $O$, one may distinguish two kinds of spatial 
relations: those {\it internal} to $O$, which hold between its material 
constituents, and those {\it external} to $O$, which hold between $O$ or its 
material constituents and objects having no constituents in common with $O$. 
By definition, a fundamental particle lacks internal spatial relations. In the 
minds of most physicists an object lacking internal spatial relations is pointlike, 
but such an object may just as well be formless. It must be emphasized that 
there can be no direct evidence of either the 
formlessness or the pointlike form of a fundamental particle.\footnote{%
There is indirect evidence: As is shown in this section, the formlessness is entailed 
by the relationist conception of space, which in turn is entailed by the behavior of 
electrons in two-slit experiments, as I shall argue in Sec.~5.}
What can be 
experimentally ascertained about a specific type of particle is the absence of 
evidence of internal structure. If absence of evidence is interpreted as evidence 
of absence -- the standard model does this in respect of the quarks and the 
leptons,~-- this may be construed either as the possession of a pointlike form 
or as the nonexistence of any form. On the latter view, the forms of all the beasts 
and baubles in this world resolve themselves into the spatial relations that 
obtain among their constituent parts. Ultimate parts have no parts, so they also 
have no form. On the former view there exists, in addition to the forms that 
resolve themselves into spatial relations, another type of spatial form, namely 
the pointlike form of a fundamental particle. It is obvious which is the more 
parsimonious view.

There are further reasons for rejecting the notion that a fundamental particle 
has a pointlike form, besides the empirical inaccessibility of such a form and 
Occam's principle of theoretical parsimony. Since space is not a storehouse of 
preexistent positions, positions need to be realized or brought into being. They 
may be realized individually -- by the presence of a material object with 
(ideally) the form of a point -- or they may be realized in pairs, as relative 
positions. But if they are realized individually, we still need to specify 
quantitatively how they are related to each other. A single object or locally 
instantiated property can define ``here'' qualitatively, but it is not enough to 
quantify (attribute a numerical value to) ``here''. For this purpose we need to 
know how ``here'' is quantitatively related to ``elsewhere''. And this is all we 
need to know. For the quantitative purposes of physics, the existence of 
position-marking forms is irrelevant. Why then should Nature go to the trouble of 
investing the ultimate spatial relata with (pointlike) forms, when it is enough to realize 
positions {\it in pairs}, as spatial relations? The fact that in the phenomenal 
world we do not and cannot encounter formless entities is no argument. As we 
shall see in Sec.~10, this is a consequence of the neurophysiological basis of the 
phenomenal world, and thus has no bearing on the nature of the physical world.

Attributing to a fundamental particle the form of a point is not only gratuitous 
but also inconsistent with the relationist conception of space. To see this, 
consider a world that contains a single fundamental particle. If this had a 
pointlike form, then there would exist, in addition to the particle, a surrounding 
spatial expanse. But if the spatial aspect of the world is a family of spatial 
relations between material objects, all that is spatially extended is the spatial 
relations between material objects. A world that contains a single fundamental 
particle contains no spatial relations, and therefore it contains nothing that is 
spatially extended. There is no surrounding spatial expanse. And so there is no 
pointlike form either, as this implies the existence of such an expanse.

I think it is high time that we recognize the notion that a fundamental particle 
has a (pointlike) form for what it is -- an unwarranted importation from the 
phenomenal world into the world of physics. In the phenomenal world, the 
existence of spatial relations 
presupposes the existence of spatially related forms (visual percepts or 
images) to which the spatial relations are attributable. Forms are 
phenomenologically prior to spatial relations. In the physical world the 
converse is true. Spatial relations are ontologically prior to forms. The form of 
an object $O$ is the set of $O$'s internal spatial relations. A noncomposite 
object lacks internal relations. Hence ``the form of a noncomposite object'' is a 
contradiction in terms.

In the days when an atom was still largely thought of as a miniature solar 
system, Werner Heisenberg, if I remember rightly, argued that if atoms are to 
explain what the phenomenal world looks like, they cannot look like anything in 
the phenomenal world -- an insight that we haven't fully assimilated yet. In the 
phenomenal world, in which objects are bundles of qualia, every object 
necessarily has a form. If the same were true of the physical world, a 
noncomposite object would have to be pointlike. But the stuff that the physical 
world is made of is very different from qualia. The physical world is made of 
formless particles and the spatial relations that hold among them.

Let us complete the list of points made in this section:

\begin{itemize}
\item Spatial forms 
are sets of spatial relations. A composite object possesses both a position 
(consisting in its external spatial relations) and a form (consisting of its internal 
spatial relations). An object that lacks internal spatial relations also lacks a 
form.
\end{itemize}

\section{The Identity of Fundamental Particles}

In what way does a material object $O$ differ from its form? What does $O$ 
possess in excess of the spatial relations that make up its form? The obvious 
answer is, the formless entities to which we attribute those spatial relations. 
And what is it that these entities contribute to $O$ over and above their spatial 
relations? The obvious answer is, existence pure and simple. The difference 
between a set $F$ of spatial relations and a material object $O$ whose internal 
spatial relations make up $F$ is that $O$ exists while $F$ by itself does not. 
What $O$ has in excess of its form is the existence that its formless material 
constituents bestow on its internal spatial relations.

An actually existing composite object $A$ possesses other properties besides 
the property of existence. A composite object therefore is something to which 
existence is contingently attributable; an object with the properties of $A$ may 
or may not exist. A relative position is likewise definable independently of its 
existence, so it, too, may or may not exist. A fundamental particle, on the other 
hand, is not something to which existence is contingently attributable. Divested 
of its external spatial relations and the dynamical parameters that contribute to 
determine the evolution of these relations,\footnote{%
Fundamental particles are usually thought of as possessing ``by themselves'' 
a mass, a spin, and various types of charge. These properties, however, are 
more appropriately understood as dynamical parameters characteristic of the 
evolution of spatial relations, rather than as intrinsic properties of individual 
particles. They tell us nothing about what a particle is in itself, or 
how it behaves out of relation to other particles.}
a fundamental particle has nothing but the property of existence. Hence it 
cannot be defined independently of its existence. If we abstract from external 
relations and dynamical parameters, there is nothing left to which existence is 
contingently attributable. All that is left is ``existence itself''. A fundamental 
particle, therefore, either is ``existence itself'' or is an instance of ``existence 
itself''.

So which is it -- ``existence itself'' or an instance of it? The answer is an 
unequivocal ``both''. To arrive at this conclusion, we will have to take account 
of another empirical finding. Let us consider a scattering experiment with 
particles of the same type -- say, two protons. Let us assume that there are 
two incoming particles, one ($N$) moving northwards and one ($S$) moving 
southwards, and two outgoing particles, one ($E$) moving eastwards and one 
($W$) moving westwards. Anyone unfamiliar with quantum mechanics will 
expect the following to be the case: either $N$ is the same particle as $W$ and 
$S$ is the same particle as $E$, or $N$ is the same particle as $E$ and $S$ is 
the same particle as $W$. Yet nothing could be further from the truth. 
Quantum mechanics tells us in no uncertain terms that neither of the incoming 
particles is identical with either of the outgoing particles.

\begin{figure}
\begin{center}
\epsfig{file=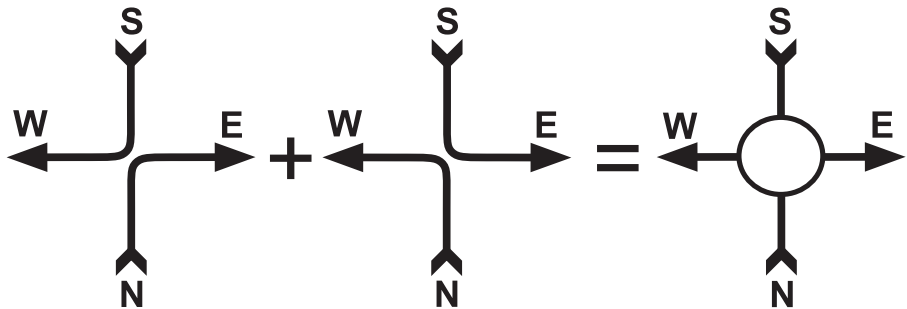,width=12cm}
\end{center}
\caption{Scattering of two particles at right angles. If the alternative processes 
on the left-hand side of this symbolic equation are distinguishable experimentally, 
the equation holds for probabilities. In this case the diagram on the right-hand side 
gives an incomplete picture of what actually happens. If the alternative processes are 
indistinguishable experimentally, the equation holds for amplitudes. In this case the 
diagram on the right-hand side gives the complete picture of what actually happens, 
while the diagrams on the left-hand side are overcomplete: They involve a distinction 
that Nature doesn't make.}
\end{figure}

One arrives at this conclusion by considering the probability $p\,(E,W)$ of the 
assumed final state (one eastbound particle and one westbound particle), 
given the assumed initial state. If the particles are of different types -- say, a 
proton and a neutron in, a proton and a neutron out -- $p\,(E,W)$ is the sum of 
two probabilities, the probability that $N$ is the same particle as $E$ and $S$ 
is the same particle as $W$, and the probability that $N$ is the same particle 
as $W$ and $S$ is the same particle as $E$, in agreement with the dictates of 
common sense:
\[
p_c(E,W)=\absosq{\braket{EW}{NS}}+\absosq{\braket{WE}{NS}}.
\]
$\braket{EW}{NS}$ and $\braket{WE}{NS}$ are the respective probability 
amplitudes associated with the alternatives ($N\rightarrow E$, $S\rightarrow 
W$) and ($N\rightarrow W$, $S\rightarrow E$), in obvious notation. The two 
possibilities that contribute to $p\,(E,W)$ are illustrated in Fig.~1. If the particles 
are of the same type, the probability for scattering at right angles is given by
\[
p\,(E,W)=\absosq{\braket{EW}{NS}+\braket{WE}{NS}}.
\]
For bosons $\braket{EW}{NS}=+\braket{WE}{NS}$, and $p\,(E,W)$ is twice as 
large as $p_c(E,W)$:
\[
p_b(E,W)=\absosq{2\braket{EW}{NS}}=4\absosq{\braket{EW}{NS}}=2p_c(E,W).
\]
For fermions $\braket{EW}{NS}=-\braket{WE}{NS}$, and $p_f(E,W)=0$. Both 
results are inconsistent with the notion that the two particles possess 
permanent identities. {\it A fortiori} they are inconsistent with the idea that a 
particle possesses the property of ``being this very particle'', known to 
philosophers as ``thisness'' or ``haecceity''.

Before and after the scattering, the two particles possess distinguishing 
characteristics: They travel in opposite directions, and they are in different 
places (relative to the laboratory frame). But at the time of scattering no such 
distinguishing characteristics exist, nor is it possible to causally link a particular 
incoming particle to a particular outgoing particle, nor can there be anything 
(such as thisness) that makes up for the missing causal identifiers. All of this 
follows from the experimentally confirmed scattering probabilities predicted by 
quantum mechanics. How many things, then, exist at the time of scattering? 
Two absolutely indistinguishable things? Or a single thing with the capacity for 
twofold instantiation?

According to the Identity of Indiscernibles, there cannot be two absolutely 
indistinguishable things. Seen from the laboratory frame, the two particles are 
absolutely indistinguishable at the time of scattering. In particular, their 
positions relative to the laboratory frame are identical.\footnote{%
If particles were impenetrable bits of stuff, it would be impossible for them to 
have identical positions. But formless entities obviously can have identical 
positions relative to a reference object or frame, and this is equally true of 
objects that are made up of formless entities.}
In spite of this, however, the two particles remain in possession of a nontrivial 
relative position, and this is sufficient for them to be two things even at the time 
of scattering. By a ``nontrivial relative position'' I mean a relative position that 
is not equivalent to an exactly vanishing distance. (Owing to the fuzziness of all 
possessed relative positions, no relative position is trivial in this sense.) The existence 
of a relation 
implies the existence of {\it two} relata. The spatial relation that exists between 
the two particles thus warrants their twoness without making them discernible 
in the laboratory frame. This permits either of two conclusions. If we consider 
the possession of distinguishing properties relative to the laboratory frame 
necessary for the discernibility of the two particles, then at the time of 
scattering there exist two indiscernible particles, and the Identity of 
Indiscernibles fails. If on the other hand we consider the existence of a not 
exactly vanishing distance between the two particles sufficient for their being 
discernible, then all particles are discernible, and the principle holds.

In either case, the reason why the two particles at the time of scattering are two 
things is their nontrivial spatial relation. Hence intrinsically (that is, considered 
{\it out of relation} to each other) 
the two particles are not two things. They are identical, and this not in the weak 
sense of exact similarity but in the strong sense of {\it numerical identity}. 
Hence the unequivocal ``both'': Intrinsically (out of relation to each other) either 
particle is ``existence itself'', and extrinsically (that is, by virtue of the spatial 
relation between them), they are two instances of ``existence itself''.

\begin{figure}
\begin{center}
\epsfig{file=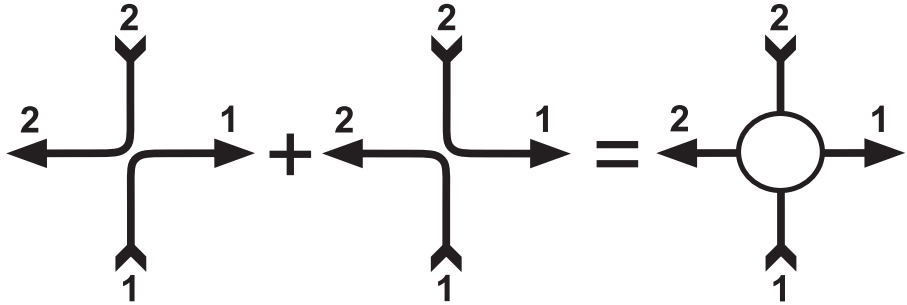,width=12cm}
\end{center}
\caption{Same as Fig. 1, except for the possibility of type swapping. 
Since the alternative processes on the left-hand side (with and without type swapping) are 
indistinguishable experimentally, the diagram on the right-hand side gives the complete 
picture of what actually happens, while the diagrams on the left-hand side again involve a  distinction that Nature doesn't make.}
\end{figure}

What has been established so far is that particles of the same type, considered 
out of relation to each other, are numerically identical. If type conversions are 
possible, this conclusion can be extended to particles of different types. 
Suppose that the two particles in our scattering experiment are of different 
types (e.g., a proton and a neutron in, a proton and a neutron out) but that 
particles of the first type can be converted into particles of the second type (a 
proton into a neutron and vice versa). Further suppose that $N$ and $E$ are of 
type~1, and that $S$ and $W$ are of type~2. Then the probability of this 
scattering event is given by
\[
p\,(E_1,W_2)=\absosq{\braket{E_1W_2}{N_1S_2}+\braket{W_2E_1}{N_1S_2}},
\]
where the indices specify the types to which the incoming and outgoing 
particles belong. Once again it is impossible to say whether a particular 
incoming particle is the same as or different from a particular outgoing particle. 
And the reason this is so is not that we have no means of knowing which of the 
alternatives depicted in Fig.~2 represents what actually happens. The reason 
this is so is that the distinction we make between the alternatives is a 
distinction that Nature does not make. Nothing in the physical world 
corresponds to the conceptual difference between these alternatives. It would 
therefore be incorrect to interpret the specified scattering event by affirming 
that the particles scatter {\it both} with and without type swapping, and it would 
be equally incorrect to interpret it by affirming that the particles scatter {\it 
neither} with {\it nor} without type swapping. Both interpretations involve a 
conceptual distinction that Nature does not make. That distinction exists solely 
in our minds.\footnote{%
As my aim here is to make sense of {\it standard} quantum mechanics, I take 
the nonexistence of hidden variables for granted. This warrants the logical step 
from a set of interfering alternatives to the objective unreality of whatever it is 
that renders the alternatives distinct to our minds.}

The same argument that took us from $p\,(E,W)$ to the numerical identity of 
particles of the same type (considered out of relation to each other), now takes 
us from $p\,(E_1,W_2)$ to the numerical identity of all particles of the same {\it 
basic} type (considered out of relation to each other). What is characteristic of a 
basic particle species is that its members cannot be converted into members of 
a different basic species. How many basic types of particle exist depends on 
the theory. According to the standard model, a member of one of the two 
species of particles known respectively as hadrons and leptons cannot be 
converted into a member of the other species (the same applies to bosons and 
fermions), while in the so-called grand unified theories hadrons and leptons 
are mutually convertible, and in supersymmetric theories ``once a fermion, 
always a fermion'' is no longer true either. In these theories, being a hadron, 
lepton, boson, or fermion is an accidental property of something that by itself is 
neither hadron nor lepton nor boson nor fermion; there exists just one basic 
type of particle. But whether or not the final theory (assuming that there will be 
one) permits conversions between all particle types, the property of belonging 
to a particular type of particle can be thought of as accidental or contingent, 
and all existing fundamental particles can be thought of as 
being intrinsically one and the same entity -- ``existence itself''.

Moreover, as was pointed out in note 8, particle species are distinguished by what 
is essentially a set of dynamical parameters governing the evolution of spatial relations. 
As attributes of an individual particle, considered out of relation to other particles, 
these parameters are meaningless. 
Hence if we consider a fundamental particle as it is in itself, we 
must mentally strip it not only of its spatial relations but also of the type 
to which it belongs. This too entails that intrinsically all fundamental particles 
are identical in the strongest possible sense: But for their spatial relations, they 
are one and the same entity. The basic reality of matter, accordingly, is not a multitude of 
fundamental particles but a {\it single entity}, and this entity cannot be qualified 
as anything more particular or determinate than ``existence itself''. Henceforth I 
shall omit the coy quotation marks and represent this entity by the symbol $\cal 
E$.

A fundamental particle is not just {\it like} $\cal E$; each fundamental particle 
{\it is} $\cal E$. At bottom the world has exactly one fundamental material 
constituent, namely $\cal E$. The individuation or multiple instantiation of $\cal 
E$ is a consequence of the realization, or the coming into existence, of spatial 
relations. $\cal E$ has the capacity to enter into spatial relations with itself, and 
this is what gives it the aspect of a multitude of relata. But what exists at either 
end of each spatial relation, considered in itself, is identically the same entity 
$\cal E$. At bottom all there is is $\cal E$ and spatial relations between $\cal 
E$ and itself. The spatial relations owe their existence to $\cal E$ -- they exist 
because they are relational determinations of $\cal E$,~-- and $\cal E$ owes its 
individuation or multiple instantiation to the existence of spatial relations: A 
multiplicity of relations implies a multiplicity of relata, even if intrinsically each 
relatum -- each fundamental particle of matter -- is $\cal E$.

Instantiation is traditionally conceived as running parallel to predication: What 
gets instantiated is a predicable universal; the resulting instance is an 
impredicable individual. This way of thinking suggests that what is responsible 
for the instantiation is something that is present in the individual but absent 
from the universal, and this idea is at the root of the Platonic-Aristotelian 
dualism of Matter and Form and its subsequent transformations, including the 
idea that physical qualities are instantiated by the ``points of space''. The 
individuation that takes us from $\cal E$ to the fundamental particles of matter 
is something else altogether. A fundamental particle is not two things -- 
(i)~instantiated existence and (ii)~something (such as a ``point of space'' or 
a part of Plato's matter-space) that does the instantiating. In a fundamental particle 
there isn't anything that is distinct from $\cal E$ and of which existence is 
predicable. There is nothing that acquires ``$\cal E$-ness'' the way a bounded 
portion of Platonic-Aristotelian matter acquires Being or actuality. There is nothing present 
in a fundamental particle that is absent from $\cal E$.

Only existence can instantiate existence, for an effective 
instantiator must exist in advance of the instantiation, and the only ``thing'' that 
exists in advance of the instantiation of $\cal E$ is $\cal E$. But this is the same as 
saying that the only way to instantiate existence is to relate it to itself. Only in this 
way can the proper logical dependences be 
implemented: The instances of existence exist because the instantiating spatial 
relations exist; the instantiating relations exist because they are properties of $\cal E$ 
(or because $\cal E$ has assumed them, or because $\cal E$ has entered into spatial 
relations with itself); and $\cal E$ exists because it is ``the one independent 
reality of which all things are an expression'' (a dictionary definition of ``the 
absolute''~\cite{Audi1995}).

The relationship between $\cal E$ and its instances thus is as close and as 
intelligible as can be. It is not some mystical relation between formless Matter 
and immaterial Form. It is identity plain and simple. The spatial multiplicity of 
the world is not based on something that is intrinsically multiple like Plato's 
matter or a set of point-individuals with the cardinality of the reals. It is a 
multiplicity of relations that entails a multiplicity of relata. But only the relata {\it 
qua} relata are many. Intrinsically (out of relation to other particles) each relatum is $\cal 
E$. The ``expression'' of the one independent reality $\cal E$ is effected by 
means of relations between $\cal E$ and itself. Physical properties are 
relational -- they are either spatial relations or dynamical parameters governing 
the evolution of spatial relations,~-- and since existence is contingently 
attributable to them, they may be thought of as universals. But $\cal E$ cannot 
be thought of as another universal. Material things are made of relations, and 
these owe their existence not to a predicable universal but to the existence that 
they relate.

\section{The Heart of Quantum Mechanics}

In the previous section we addressed the ``the miraculous identity of particles 
of the same type'', which Misner et al.~\cite[p.~1215]{Misneretal1973} regard 
as ``a central mystery of physics''. In this section we examine a phenomenon 
which according to Feynman et al.~\cite[Sec.~1--1]{Feynmanetal1965} ``has in 
it the heart of quantum mechanics''.

Let $R$ be some bounded spatial region. We tend to think that the world can 
be divided into things or parts that are inside $R$ and things or parts that are 
outside $R$. We tend to think that spatial distinctions like the distinction 
between the inside of $R$ and the outside of $R$ are real {\it per se}, and that {\it a 
fortiori} they are real for everything that exists in space. If this were the 
correct way of thinking about space, any object $O$ would at any time satisfy 
one of the following three conditions: (i)~$O$ is wholly inside $R$; (ii)~$O$ is 
wholly outside $R$; (iii)~$O$ has two parts, one inside $R$ and one outside 
$R$. Once again reality does not comply with the dictates of common sense. 
Quantum mechanics tells us in no uncertain terms that sometimes all of these 
propositions are false.

The paradigm example is a two-slit interference experiment with 
electrons~\cite[Chap.~1]{Feynmanetal1965}. 
Anyone unfamiliar with quantum mechanics expects one of the following 
propositions to be true of each electron: (i)~it goes through the first slit, (ii)~it 
goes through the second slit, (iii)~it consists of parts that go through different 
slits. Yet whenever the experimental arrangement is such that interference 
fringes are observed, the electron does none of this. It doesn't go through a 
particular slit,\footnote{%
There is nothing so obvious that a philosopher cannot be found to deny it, as 
Lockwood~\cite[p.~45]{Lockwood1989} observed. It seems to be the same with 
physicists. Bohmian mechanics~\cite{Bohm1952} tries to salvage as much of 
common sense as is consistent with the empirical data, by attributing to each particle 
a definite but observationally meaningless position. By introducing 
unobservable (and hence arbitrary) causes for stochastic events, this 
theory makes each electron go through a definite slit.}
and it doesn't get divided by its passage through the 
slits. Albert~\cite[p.~11]{Albert1992} draws from this the conclusion that 
``[e]lectrons seem to have modes of being, or modes of moving, available 
to them which are quite unlike what we know how to think about''.

The behavior of electrons becomes intelligible if we reject the substantival,
set-theoretic conception of physical space. The three propositions just 
considered are based on the assumption that the two slits are distinct {\it per 
se}. One of the things that the interference fringes are trying to tell us is that 
our conceptual distinction between the regions defined by the slits does not 
exist for the electron. Hence the distinction cannot be real {\it per se}. In other 
words, it cannot be intrinsic to space. Hence the substantival, set-theoretic 
conception of physical space, which entails the distinction, cannot be the right 
way of thinking about physical space.

The electron is able to go through the two slits {\it indiscriminately} (that is, 
without going through a particular slit and without being divided) because the 
conceptual distinction between ``disjoint parts of space'' (which is inherent in 
the set-theoretic conception of space) does not always correspond to 
something that is physically real. When it is appropriate to add amplitudes 
rather than probabilities,\footnote{%
As will become clear in the next section, this is the case when the following two 
conditions are fulfilled: (i)~There isn't any matter of fact about the alternative 
taken, and (ii)~the quantum system under consideration, $S_1$, is not 
correlated (``entangled'') with another quantum system $S_2$ in such a way 
that inferences to the alternative taken by $S_1$ could be drawn from future 
measurements performed on $S_2$.}
the distinction we make between the corresponding alternatives is a distinction 
that Nature does not make. We can say that the electron went through both 
slits if ``both slits'' stands for a single positional attribute -- the opening made 
up of the two slits. But we cannot make this equivalent to two propositions 
(``the electron went through the first slit'' and ``the electron went through the 
second slit'') since this involves a distinction that has no counterpart in the 
physical world. The distinction exists solely in our minds. But if a distinction 
exists solely in our minds, the same is true of any notion that implies the 
distinction. The notion which makes the two alternatives in a two-slit experiment 
distinct for us, and the behavior of electrons incomprehensible, is the notion 
that physical space is intrinsically and infinitely differentiated. The notion that 
the multiplicity of \R\ is inherent in physical space (and that, consequently, the 
individual points of space or space-time can be regarded as carriers of physical 
properties) is as inconsistent with quantum mechanics as the notion of 
absolute simultaneity is with special relativity. This notion perhaps more than 
any other is what prevents us from making sense of quantum mechanics.

If space were an intrinsically and infinitely differentiated constituent of the 
physical world, then any spatial region, however small, would {\it per se} be 
distinct from any other (disjoint) spatial region, and every material object $O$ 
would consist of as many spatial parts as the space it occupies. The parts of 
space would {\it define} the parts of $O$. If $O$ has only a finite number of 
material constituents, these would necessarily be pointlike, and for any 
partition $\{d^3R\}$ of space into infinitesimal regions the following would 
necessarily be the case: Each of the material constituents of $O$ is located 
inside a particular region $d^3R$. In other words, on a substantival, set-%
theoretic account of physical space, the positions of things are necessarily 
definite. In reality -- so the behavior of the electrons tells us -- space is a 
system of spatial relations between material objects, and these spatial relations 
are (more or less) indefinite. The behavior of electrons in two-slit interference 
experiments is Nature's corroboration of the relationist conception of physical 
space developed in Secs.~2 and 3.

Such an experiment typically features several macroscopic objects like an 
electron gun (a heated tungsten wire in a metal box with a hole, the wire being 
at a negative voltage with respect to the box), a thin metal plate with two slits in 
it, and an array of detectors (Geiger counters or electron multipliers connected 
to loudspeakers). What these objects have in common is that their relative 
positions can be treated as definite.\footnote{%
Why this so is explained in Sec.~7. Sometimes an exception is made for the slit 
plate. If the transverse momentum of the slit plate is so sharp that by 
measuring it one can infer the slit taken by the electron, the positional 
indefiniteness of the slit plate (relative to the rest of the apparatus) rules out 
the existence of an interference pattern~\cite[Sec.~1--8]{Feynmanetal1965}. 
The correct explanation of the obliteration of the interference pattern, however, 
is not the positional indefiniteness of the slit plate but the ensuing {\it 
correlation} of the transverse momentum of the slit plate with the transverse 
position of the electron.}
The only objects with pertinently indefinite positions (relative to any 
macroscopic part of the apparatus) are the electrons coming from the electron 
gun. We tend to visualize the position $\cal P$ of such an electron (relative to 
the apparatus) as a cloud or a smudge (a ``wave packet'' or, worse, two ``wave 
packets''). It is important to understand that this picture cannot be taken to 
represent a state of affairs in physical space. The ``space'' containing the cloud 
is the set $C$ of values that are available for attribution to the relative position 
$\cal P$. This value ``space'' must not be conflated with the physical space 
$S$, which contains $\cal P$. It is a set of triplets of real numbers. It does not 
possess the quality of spatial extension known to us from percepts and images 
in phenomenal space. Outside of phenomenal space, this quality is attributable 
solely to the possessed spatial relations that make up $S$.

The ``point of contact'' between $C$ and $S$ is {\it detectors}. By a ``detector'' 
I mean anything that is capable of indicating the presence of a material object, 
such as an electron, in a particular spatial region $R_i$, or its passing through 
a particular section $\sigma_i$ of a plane. The regions $R_i$ or the sections 
$\sigma_i$ are physically realized or realizable with the help of macroscopic 
boundaries (that is, boundaries made up of a large number of material 
constituents the spatial relations between which can be treated as definite). 
These regions or sections define values (``inside $R_i$'' or ``through 
$\sigma_i$'') that are potentially attributable to $\cal P$. They therefore ``exist'' 
both in $S$ and in $C$. But only if a detector clicks (that is, only if there is a 
matter of fact about the particular region containing the electron or the 
particular section crossed by the electron) is the corresponding value actually 
attributable to $\cal P$, as will become clear in what follows.

If the substantival, set-theoretic conception of physical space is inconsistent 
with quantum mechanics, so is the doctrine called local realism, according to 
which, ``\ldots\ all there is to the world is a vast mosaic of local matters of 
particular fact, just one little thing and then another\ldots. We have geometry: 
a system of external relations of spatiotemporal distance between points\ldots. 
And at those points we have local qualities: perfectly natural intrinsic properties 
which need nothing bigger than a point at which to be instantiated\ldots. And 
that is all\ldots. All else supervenes on that'' \cite[p.~X]{Lewis1986}. What 
quantum mechanics is trying to tell us is that reality is not built ``from the 
bottom up'', on an infinitely and intrinsically differentiated space, out of locally 
instantiated physical properties. There are no points on which a world of such 
properties can be built. Reality instead is built ``from the top down'': By 
assuming a multitude of spatial relations, $\cal E$ takes on not only the aspect 
of a spatially differentiated world but also the aspect of a multiplicity of 
fundamental particles.

If the doctrine of local realism is an exploded myth, so is the principle of local 
action (a.k.a. local causality), according to which the ``local matters of 
particular fact'' at a point $x$ are influenced only by the ``local matters of 
particular fact'' in the infinitesimal neighborhood of~$x$. Local action therefore 
cannot be the solution to the perceived problem of action at a distance. Nor 
does this require a solution, since it is a pseudo-problem arising from 
erroneous assumptions. For one thing, it presupposes the existence of 
separate interacting entities $A$ and $B$, as well as the existence of separate 
regions of space containing $A$ and $B$. But $A$ and $B$ aren't separate 
entities, nor are there separate regions of space. If $A$ and $B$ are 
fundamental particles, they both are $\cal E$, and if they are composite 
objects, the ultimate material constituents of $A$ are numerically identical with 
the ultimate material constituents of $B$. Physical space, on the other hand, 
isn't something that by itself has parts; it is a system of spatial relations 
between material objects. The spatial differentiation of the physical world is a 
property of its material content, not of a substantivally conceived space. Two 
locations are therefore never separate ``by themselves''. Only material objects 
can be spatially separated, and for two objects $A$ and $B$ to be spatially 
separated, it is necessary that their relative position ${\cal P}(A,B)$ is 
sufficiently large compared to its indefiniteness. But this is not enough: The 
position of $B$ relative to $A$, {\it as seen from the laboratory frame}, ${\cal 
P_L}(A,B)={\cal P}(B,L)-{\cal P}(A,L)$, must likewise be sufficiently large 
compared to its indefiniteness.\footnote{%
$L$ marks the origin of the laboratory frame. While ${\cal P}(A,B)$ corresponds 
to a single conditional probability distribution -- the distribution of ${\cal P}(B,L)$ 
conditional on $A$'s having a numerically exact position in the laboratory 
frame,~-- ${\cal P_L}(A,B)$ corresponds to the difference between the 
distributions of ${\cal P}(B,L)$ and ${\cal P}(A,L)$. (A word of 
caution: As we shall see below, the probabilities that one may use to 
quantitatively describe indefinite relative positions are distributed 
over counterfactuals. This rules out naive realistic interpretations of position 
probability distributions and wave functions, such as Schr\"odinger's original 
interpretation of $\psi$ as some bizarre kind of real jelly.)}
The following often cited statement~\cite{Einstein1948} is therefore 
unfounded: ``An essential aspect of [the] arrangement of things in physics is 
that they lay claim, at a certain time, to an existence independent of one 
another, provided these objects `are situated in different parts of space'$\,$''. 
There are no ``different parts of space''.

For another thing, the very attempt to {\it explain} (causally or otherwise) the 
fundamental behavior of matter appears misconceived. Explanation {\it begins} 
with the fundamental behavior of matter, which can only be {\it described}, and 
which in part is described by quantum mechanics. We have space, which is a 
system of relations between $\cal E$ and itself. In classical physics we have 
dynamical laws, which spell out how these relations evolve in time. In quantum 
physics we have mathematical condensations of statistical regularities. The 
idea that particles {\it act on} particles, or that the distribution and motion of 
matter ``here'' and ``now'' is {\it causally} related to the distribution and motion 
of matter ``there'' and ``then'', or that measurement results are so related, adds 
nothing of physical significance to (classical) laws that simply describe how 
spatial relations are temporally related, or to (quantum-mechanical) laws that, 
as will become clear in what follows, describe statistical correlations between 
property-indicating facts. And specifically, any attempt to explain the statistical 
regularities in terms of causal strings that criss-cross space or space-time 
involves the principle of local action and therefore is inconsistent with quantum 
mechanics.

\section{Quantum-Mechanical Properties Are Extrinsic}

Quantum mechanics is, if nothing else, a tool for calculating probabilities. It 
represents the possible values $q^k_i$ of observables $Q^k$ as projection 
operators ${\bf P}_{Q^k=q^k_i}$ on some Hilbert space $\cal H$. The 
projection operators that jointly represent the range of possible values of a 
given observable are mutually orthogonal. If one defines the ``state'' of a 
system as a {\it probability measure} on the projection operators on $\cal 
H$ resulting from a {\it preparation} of the system 
\cite{Cassinello-SG1996}\cite[p.~92--94]{Jauch1968}, one finds 
\cite{Cassinello-SG1996, Gleason1957}\cite[p.~132]{Jauch1968}
that every such probability measure has the form $p\,({\bf 
P})=\mbox{Tr}({\bf WP})$, where $\bf W$ is a unique density operator [that is, a 
unique self-adjoint, positive operator satisfying $\mbox{Tr}({\bf W})=1$ and 
${\bf W}^2\leq{\bf W}$]. $\mbox{Tr}$ signifies the 
trace defined by the formula $\mbox{Tr}({\bf X}):=\sum_i\BOK{i}{{\bf X}}{i}$ for any 
orthonormal basis $\{\ket{i}\}$. If ${\bf W}^2(t)={\bf W}(t)$, ${\bf W}(t)$ projects 
on a one-dimensional subspace of $\cal H$ and thus is equivalent -- apart from 
an irrelevant phase factor -- to a state vector $\ket{\psi(t)}$ or a wave function 
$\psi({\bf r},t)$, $\bf r$ being any point in the system's configuration space.

It is tempting to attribute the truth of quantum-mechanical probability 
assignments to an underlying state of affairs. It is equally tempting to assume 
that this state of affairs is somehow represented by the state vector or the 
density operator. Instead of jumping to such conclusions, however, one should 
heed van Kampen's~\cite{vKampen1988} warning: ``Whoever endows $\psi$ 
with more meaning than is needed for computing observable phenomena is 
responsible for the consequences\ldots.'' One should not lose sight of the fact 
that the state vector or the density operator is what one gets if one {\it defines} 
the ``state'' of a system as a probability measure on the projection operators 
representing the possible properties of the system. The idea that what by 
definition is a tool for assigning probabilities to {\it possibilities} also describes 
an {\it actual} state of affairs, is simply a category mistake.

Quantum mechanics being inconsistent with local realism, it is impossible to 
interpret $\psi({\bf r},t)$ as a local quality -- something that ``need[s] nothing bigger 
than a point at which to be instantiated''. The fact that we live in a relativistic 
world, moreover, commits us to treating time and space on an equal footing, as 
far as this is consistent with the obvious qualitative differences between the 
two. Specifically, if space is a system of spatial relations, then time is a system 
of temporal relations (or, rather, then space-time is a system of spatiotemporal 
relations), and if space isn't intrinsically and infinitely differentiated (or, rather, if 
the world isn't infinitely differentiated spacewise), then time isn't intrinsically 
and infinitely differentiated either (that is, the world isn't infinitely differentiated 
timewise). In other words, if space isn't a set of self-existent points, time can't 
be a set of self-existent instants. Hence $\psi({\bf r},t)$ not only isn't something that 
exists at ``the point $\bf r$'' but also isn't something that exists at ``the instant 
$t$''. Once this is understood, there doesn't seem to be much point in 
construing the state vector as an actual state of affairs, for the principal reason 
for doing so has always been to protect the golden calf of determinism, which 
requires local realism in general and the local reality of $\psi({\bf r},t)$ in particular. 
That probabilities are nonlocal in space and in time makes perfect sense: The 
probability for an object to be found inside a region $R_i$ at a time $t$ is not 
something that exists inside the region $R_i$ or at the time $t$. A state of 
affairs that is nonlocal in time and space, on the other hand, is not something 
that we know how to make sense of.

That it makes no sense to construe the state vector as an actual state of affairs 
does not mean that one cannot make ontological sense of quantum 
mechanics. However, in order to do so one must steer clear of two diametrically 
opposed mistakes. The first is the error of the instrumentalist, who regards as 
idle metaphysics any attempt to give a realistic or epistemological account of 
how it is that the statistical regularities predicted by quantum mechanics come 
out the way they do. For the instrumentalist, the results of measurements are 
not possessed properties, for there are no ``quantum objects'' that could 
possess them; there is no ``quantum world'' to which the results of 
measurements could be attributed.

The instrumentalist throws out the baby with the bath water. Suppose that we 
perform a series of position measurements, and that every position 
measurement yields exactly one result (that is, each time exactly one detector 
clicks). In this case we are entitled to infer the existence of an entity $O$ that 
persists through time, to think of the clicks given off by the detectors as matters 
of fact about the successive positions of this entity, to think of the behavior of 
the detectors as position measurements, and to think of the detectors as 
detectors. The case for instrumentalism is that the position-indicating clicks are 
not only sufficient but also necessary for the existence of the positions 
indicated by the clicks. That is why Bohr~\cite{Bohr1934, Bohr1958} insisted on the 
necessity of describing quantum phenomena in terms of the experimental 
arrangements in which they are displayed. But this does not mean that there is 
no object to which the indicated positions can be attributed. What it means is 
that the contingent properties\footnote{%
A property $q$ attributable to $O$ is a {\it contingent} property of $O$ iff it is 
not necessarily possessed by $O$.}
of the quantum world are {\it extrinsic} rather than intrinsic, in the sense that the 
existence of a fact 
indicating that $O$ is (has the property) $q$ is necessary for $O$'s being $q$: 
The properties of the quantum world {\it are} what can be inferred from 
property-indicating facts (including the outcomes of laboratory experiments) or 
from measurement results (if ``measurement result'' is taken in the general 
sense of ``property-indicating fact''). This tells us something significant about 
the quantum world, rather than controverts the existence of a quantum world.

On this interpretation, the number of objects that exist at any one time is as 
extrinsic as are the properties that can be attributed to them. It is only because 
every time the same number of detectors click (in the previous example, one) 
that there exists, at the times of the position measurements, the same 
determinate number of objects. The quantum world lacks intrinsic properties, 
including the property of containing a determinate number of objects. But, 
again, this is not the same as saying that there is no quantum world. There is 
something to which a particular number of spatial relations and a particular 
number of corresponding relata can be attributed, provided that the attribution 
is warranted by facts. This something is $\cal E$. The quantum world is 
$\cal E$ plus whichever properties are warranted by the facts, including the 
number of times that $\cal E$ is instantiated.

The opposite mistake is the error of the quantum realist, who endorses the 
following ``sufficiency condition''~\cite[p.~72]{Redhead1987}: ``If we can predict 
with certainty, or at any rate with probability one, the result of measuring a 
physical quantity at time~$t$, then at the time $t$ there exists an element of 
reality corresponding to the physical quantity and having a value equal to the 
predicted measurement result''. To begin with, $p\,(R_i,t)=1$ 
does not mean that the probability of finding $O$ in $R_i$ is one at the 
time~$t$. Instead, it means that the probability of finding $O$ in $R_i$ at the 
time $t$ is one. There is a significant difference between these two 
interpretations. Since probabilities are not things that exist in time or at times, 
one cannot speak of a probability as {\it having} such and such a value {\it at} 
such and such a time. One can only speak of the probability of finding $O$ 
in $R_i$ at the time~$t$. This probability exists neither inside $R_i$ nor at 
the time~$t$. The insufficiency of Redhead's ``sufficiency condition'' hinges on 
the nonexistence of the particular time $t$ in the absence of an actual 
measurement performed at the time~$t$, as I proceed to show.

A particular region $R$, recall, does not exist unless it is physically realized 
with the help of a macroscopic boundary, as defined in the previous section. 
And it does not exist for $O$ (that is, the difference between ``$O$ is inside 
$R$ at $t$'' and ``$O$ is outside $R$ at $t$'' has no physical reality) unless 
there is a matter of fact about $O$'s whereabouts with respect to $R$ (``inside'' 
or ``outside'') at the time~$t$ (that is, unless there is an actual event or state of 
affairs, such as the click of a detector, from which the truth of either ``inside'' or 
``outside'' can be inferred). Since the spatial distinctions that we make in our 
heads are not physically real {\it per se}, they need to be realized in order to be 
physically real -- they need a physical ``hook'' from which they can ``dangle''. 
This hook is to be sought among the actual events and states of affairs that 
happen or obtain, and specifically among those that indicate the value of a 
relative position (from a given range of values such as ``inside $R_i$'', 
$i=1,\ldots,n$). The possessed values of relative positions dangle from 
position-indicating facts. And so do the actually existing spatial distinctions, 
since they are contingent on the possessed values of relative positions.

An immediate consequence of the extrinsic nature of contingent properties is 
that the position of one object $A$ relative to the laboratory frame may be 
definite with respect to a given partition of this frame (there may be a matter of 
fact about the particular region containing $A$), while the position of another 
object $B$ may be indefinite with respect to the same partition (there may not 
be any matter of fact indicating the particular region containing $B$). In other 
words, a particular spatial distinction may be real for one object and not real for 
another. The particular region $R$ may exist for one object and not exist for 
another.

The particular time $t$, likewise, does not exist unless it is physically realized. 
Since temporal distinctions are not real {\it per se}, they too need to be 
realized; they too need a physical hook from which they can dangle. Once 
again the hook is to be found among the actual events and states of affairs that 
happen or obtain, but now specifically among those that indicate not only the 
possessed value of some observable but also the time at which this is 
possessed. The times at which the extrinsic properties of the quantum world 
are possessed are themselves extrinsic: They dangle from facts indicating both 
{\it that} a property is possessed and {\it when} it is possessed. An immediate 
consequence of the extrinsic nature of the times at which properties are 
possessed, is that a particular time $t$ may be real for one object and not real 
for another object. For instance, it may be real for a laboratory clock but not 
real for the quantum system at hand. Where $O$ is concerned, the particular 
time $t$ exists iff the possession by $O$ of some property $q$ at the time $t$ is 
factually warranted (that is, iff there is a property $q$ such that the inference 
``$O$ is $q$ at $t$'' is warranted by facts). Otherwise $t$ does not exist for 
$O$, and $O$ cannot be said to exist at the particular time $t$.

In nonrelativistic quantum mechanics, where the number of objects or 
subsystems is constant, we feel that we may imagine a particle as existing 
also between the times at which it has factually warranted properties, and we 
feel that we may imagine it as permanently possessing such invariable 
properties as a mass or an electric charge. If a region $R$ is surrounded by an 
impenetrable barrier, and if the presence of $O$ in $R$ at a time $t$ is 
factually warranted, we further feel that we may imagine $O$ as being inside 
$R$ also before and after the time $t$. But let us be clear about the meanings 
of the temporal referents ``between the times at which it has factually 
warranted properties'' and ``before and after the time $t$''. Since time is not a 
set of instants, this cannot mean ``at any or every instant between those times'' 
or ``at any or every instant before or after $t$''. It can only mean ``during the 
continuous, undifferentiated time spans between those times'' or ``during the 
continuous, undifferentiated time spans before and after $t$''. Are these 
legitimate temporal referents?

A particular time $t'$ exists only if it is indicated by an actual event or state of 
affairs, and it exists for $O$ only if it is the factually warranted time of 
possession, by $O$, of a factually warranted property. Are the undifferentiated 
time spans $\tau_i$ between the particular times $t_i$, at which $O$ has 
factually warranted properties $q_i$, nevertheless legitimate temporal referents 
for attributions to $O$? The answer depends on whether the facts that indicate 
the possession of the respective properties $q_i$ at the respective times $t_i$ 
can be thought of as indicating the possession of the same properties during the time 
spans $\tau_i$. In nonrelativistic quantum mechanics, this is the case where 
the property of individuated existence is concerned: If the existence of an individual object 
$O$ at the times $t_i$ is warranted by actual events or states of affairs (that is, if $O$ is 
detected at the times $t_i$) then $O$ also exists as an individual during the 
intervening time spans $\tau_i$. And if the presence of $O$ in $R$ at the time 
$t$ is warranted by some actual state of affairs, this state of affairs also 
warrants the presence of $O$ in $R$ during the preceding and succeeding 
undifferentiated time spans -- but {\it not} at any particular time $t'$ during 
these time spans. Unless the presence of $O$ in $R$ (or the possession by 
$O$ of any other contingent property) at $t'$ is factually warranted, the 
particular time $t'$ does not exist for $O$, and $O$ cannot be said to exist at $t'$.

Thus we can predict (as well as retrodict) with probability one that $O$ will be 
(would have been) found inside $R$ at the time $t'$ if the appropriate 
measurement is (had been) performed at $t'$, but we cannot affirm that at the 
particular time $t'$ there exists an element of reality corresponding to the 
presence of $O$ in~$R$. Unless the measurement is actually performed 
at~$t'$, $t'$~does not exist for $O$, so an element of reality involving $O$ 
cannot exist at $t'$. While the instrumentalist errs by ignoring the possibility of 
a quantum world with extrinsic properties, the quantum realist errs by ignoring 
the fact that the properties of the quantum world are extrinsic, and that, 
therefore, probability one is not sufficient for the existence of an element of 
reality.\footnote{%
An anonymous referee (of a different paper and a different journal) asserts that 
standard quantum mechanics does not include Redhead's ``sufficiency 
condition'' but instead encompasses the ``eigenstate-eigenvalue link'', 
according to which an element of reality corresponding to an eigenvalue of an 
observable exists at time $t$ iff the system at $t$ is ``in the corresponding 
eigenstate of this observable''. This too is incorrect, and for exactly that same 
reason.}

In a relativistic world, in which the number of objects or subsystems is itself a 
variable and contingent property, we may imagine a closed system as existing also 
between the times at which it has factually warranted properties, and we may 
attribute to it its conserved quantities during these undifferentiated time spans, 
provided that at some time their possession is warranted by facts. But 
between the times at which the number of subsystems has a factually 
warranted value, there is only one logical subject that has a counterpart in the 
physical world, namely existence itself. Apart from the conserved quantities 
that are enduringly attributable to it, this has properties (including the number 
of times that it is instantiated) only if and only when they are 
warranted by facts.

\section{Indefiniteness, Macroscopic Objects,
\\And the Emergence of Causality}

The conceptual innovation due to the Copenhagen interpretation of quantum 
mechanics (CIQM) has been characterized by Stapp~\cite{Stapp1972} in the 
following words: ``The theoretical structure did not extend down and anchor 
itself on fundamental microscopic space-time realities. Instead it turned back 
and anchored itself in the concrete sense realities that form the basis of social 
life.'' The first part of this statement belongs to the core of the CIQM, and it is 
what makes this interpretation superior to any interpretation that allows for 
fundamental microscopic space-time realities. There is no general consensus 
as to the other claims that form part of the CIQM. Bohr's cryptic and not always 
consistent utterances on the subject have been invoked to support a variety of 
conflicting readings. Stapp's reading restricts physics to perceived and 
communicable phenomena. Science, however, is driven by the desire to know 
how things {\it really} are, and it owes its immense success in large measure to 
the belief that this can be discovered. Unless there is conclusive proof to the 
contrary, it would be premature to relinquish this powerful ``sustaining 
myth''~\cite{Mermin1996}. But if the theoretical structure is anchored neither in 
microscopic space-time realities nor in concrete sense realities, then what 
supports it? The answer is, property- and time-indicating facts.

Loewer~\cite{Loewer1998} associates with the CIQM the following claim: ``An 
isolated quantum system evolves in conformity with a linear deterministic law 
(Schr\"odinger's equation) unless it is measured. Measurements are governed 
by an indeterministic law -- the collapse postulate.'' This -- the von
Neumann-Dirac interpretation -- cannot be part of the CIQM, given that the 
core of the CIQM rejects elementary space-time realities and therefore rejects 
elementary time realities. Because time is not a set of such realities (instants), 
a measurement does not prepare an instantaneous state of affairs that crawls 
predictably through an intrinsically and infinitely differentiated time until it 
unpredictably changes into a different instantaneous state of affairs. All that a 
measurement ``prepares'' is probabilities, and probabilities are not things that 
exist or evolve in time. Nor are density operators and state vectors such things, 
considering that they themselves are essentially (that is, by definition) 
probability measures on the possible outcomes of measurements. ``[T]here is 
no interpolating wave function giving the `state of the system' between 
measurements''~\cite{Peres1984}.

Quantum-mechanical probabilities are conditional. The (standard) 
Born probabilities are conditional on (i)~a prediction basis (the actual events or 
states of affairs that determine the ``preparation''), (ii)~the observable $Q$ 
that is being measured (including its range of possible values), (iii)~the time 
$t$ at which this is measured, and (iv)~the existence of a measurement result 
(that is, a matter of fact about the actual value of $Q$ at the time $t$). ABL 
probabilities, named after Aharonov, Bergmann, and Lebowitz~\cite{ABL1964}, 
are conditional on an inference basis rather than a prediction basis. This 
inference basis includes, in addition to the ``preparation'', the actual events or 
states of affairs that determine the ``retroparation'' of the 
system~\cite{Aharonov-Vaidman1991, Reznik-Aharonov1995, Vaidman1998, 
Mohrhoff2000}.\footnote{%
The ABL probability with which a measurement of the observable $Q$ between 
the ``preparation'' represented by the state $\ket{\psi_1}$ and the 
``retroparation'' represented by the state $\ket{\psi_2}$ yields the result $q_i$, 
is given by the ABL formula
\[
P_{ABL}(q_i)={\absosq{\sandwich{\psi_2}{{\bf P}_{Q=q_i}}{\psi_1}}\over
\Sigma_j\absosq{\sandwich{\psi_2}{{\bf P}_{Q=q_j}}{\psi_1}}},
\]
where the ${\bf P}_{Q=q_i}$ projects on the subspace corresponding to the 
eigenvalue $q_i$ of $Q$~\cite{Aharonov-Vaidman1991}.}
Quantum-mechanical probabilities are determined by facts (not mediately via 
an evolving, collapsible, instantaneous state of affairs but directly), and they 
cannot be assigned without specifying both an observable and a 
time at which this is measured. Hence it should be clear that the time on which 
quantum-mechanical probabilities, density operators, and state vectors depend 
is {\it the specified time of a specified measurement}, rather than the time of an 
evolving state of affairs. The parameter $t$ in the (Born) probability distribution
\[
p\,(R_i,t)=\mbox{Tr}\bigl({\bf W}(t)\,{\bf P}(R_i)\bigr)=\BOK{\psi(t)}{{\bf P}(R_i)}{\psi(t)}=
\int_{R_i}d^3r\,\psi^*({\bf r},t)\,\psi({\bf r},t)
\]
is not the time at which the probability of finding $O$ in $R_i$ has the value 
$p\,(R_i,t)$. Instead it is either the time of an actually performed measurement 
determining the particular region $ R_i$ containing $O$ or the specified time of 
a counterfactually performed such measurement.

Nothing in quantum physics corresponds to the intrinsically and infinitely 
differentiated time through which, according to a certain folk belief, the present 
``moves'' or ``advances''. Nothing in quantum physics corresponds to the 
related notion of an instantaneous state that ``moves'' or ``advances'' through 
an ordered set of preexistent instants having the cardinality of the ``real line''. 
And therefore nothing in quantum physics warrants the other folk belief 
according to which causal influences are carried towards the future by an 
instantaneous state of affairs (and therefore, in a relativistic world, conformably 
to the principle of local action). Where quantum physics is concerned, all the 
determining that goes on in the physical world is the determining by property-%
indicating facts (actually obtained measurement results) of {\it probabilities} 
associated with the possible results of specified measurements that are 
actually or counterfactually performed at specified times. What does no take 
place in the physical world is a determining by property-indicating facts of 
other property-indicating {\it facts}. There are no causal links between factually 
warranted properties or between the corresponding property-indicating actual 
events or states of affairs.

The probabilities of classical physics are subjective. They come into play 
whenever the exact state of a system is unknown, intractable, or irrelevant to 
the problem at hand. Quantum-mechanical probabilities, on the other hand, 
have an objective as well as a subjective application. If observable $Q$ is 
actually measured, the probabilities associated with the range of possible 
values of $Q$ are subjective; they are based on a limited knowledge that does 
not take account of the actual measurement result. This applies not only to 
Born probabilities, which take account of the system's ``preparation'' only, but 
also to ABL probabilities, which also take account of the system's 
``retroparation''. On the other hand, if no measurement is made at the time $t$, 
the ABL probabilities associated with $Q$'s range of possible values and the 
specified time $t$ are objective, in the sense that they have nothing to do with 
ignorance. All relevant facts have been taken into account; there is nothing for 
us to be ignorant of. Born probabilities in general contain a subjective element 
even if $Q$ is not actually measured, for they ignore the relevant matters of 
fact about the properties of the system at later times. Born probabilities can be 
objective only if there aren't any matters of fact about the future properties of 
the system. Thus objective probabilities are always assigned to counterfactuals 
-- conditional statements that presuppose the falsity of their antecedents, 
which are of the form ``If $Q$ were measured'',~-- and if there are relevant 
matters of fact about the future properties of the system, they have to be 
calculated using the ABL formula.\footnote{%
Kastner's~\cite{Kastner1999a,Kastner1999b} objection to 
Vaidman's~\cite{Vaidman1999a,Vaidman1999b,Vaidman1999c} counterfactual 
usage of the ABL formula has no bearing on my counterfactual usage of this 
formula, as may be gleaned from Kastner's withdrawal of a 
paper~\cite{Kastner2000} in which she raised similar objections to the way I 
assign ABL probabilities to counterfactuals in my 
forthcoming~\cite{Mohrhoff2000}, admitting that her paper was ``based on a 
misunderstanding of Mohrhoff's use of the term `counterfactual'$\,$'' and that 
``Mohrhoff's counterfactual uses of the ABL rule correspond to special cases in 
which such use is valid''. The counterfactuals to which I assign ABL 
probabilities have antecedents of the form ``if $Q$ had been measured 
between $t_1$ and $t_2$ (while actually no measurement is made between the 
preparation at $t_1$ and the retroparation at $t_2$)'', whereas it was Kastner's 
initial impression that I allow antecedents of the form ``if $Q$ had been 
measured between $t_1$ and $t_2$ (while actually a different, noncommuting 
observable $Q'$ was measured between $t_1$ and $t_2$)''. That my use of the 
ABL formula is the correct use also transpires from Cohen's~\cite{Cohen1995} 
analysis.}

Now is the time to make good on my promise to define what exactly I mean by 
an ``indefinite'' or ``fuzzy'' position. Let $\{R_i|i=1,\ldots,n\}$ be some partition of 
space (that is, of the ``space'' of values available for attribution to the positions 
of objects relative to some reference object $\cal O$.) Let $O(t)\subset R$ 
denote the proposition ``Object $O$ is inside region $R$ at time~$t$''. (This means that 
at the time $t$ the position of $O$ relative to $\cal O$ has the value ``inside $R$''; 
it does not mean that it has a definite value falling inside the range $R$.) 
Let $Q$ be the particular position observable whose range of possible values is 
$\{O(t)\subset R_i|i=1,\ldots,n\}$. Finally, let $Q\Rightarrow O(t)\subset R_k$ 
stand for the conditional ``If $Q$ is (or were) measured at the time $t$, $O$ is 
(or would be) found inside $R_k$''. The position of $O$ is {\it indefinite with 
respect to} $\{R_i|i=1,\ldots,n\}$ iff (i)~the conditionals $\{Q\Rightarrow 
O(t)\subset R_i|i=1,\ldots,n\}$ are counterfactuals ($Q$ is not actually 
measured) and (ii)~the objective probabilities associated with these 
counterfactuals are positive for at least two $i$. For instance, if there isn't any 
matter of fact concerning the slit taken by an electron, the electron's transverse 
position at the time of its passing the slit plate is indefinite just in case the 
probabilities associated with the following counterfactuals are positive: ``If there 
were a matter of fact about the slit taken by the electron, it would indicate that 
the electron went through slit $i$'' ($i=1,2$).

Note that $O$'s position at a specified time $t$ may be indefinite with respect 
to some partition or range of possible values and definite with respect to another 
partition or range of possible values. Consider, for instance, a three-slit 
experiment incorporating a device $D$ that is capable of indicating whether the 
electron went through slit $A$ or through the union $B\cup C$ of the remaining 
slits, but not capable of distinguishing between ${\bf e}_B$~= ``the electron 
went through slit $B$'' and ${\bf e}_C$. Suppose that $D$ indicates that the 
electron went through $B\cup C$. If the respective probabilities associated with 
${\bf e}_B$ and ${\bf e}_C$ are positive and objective, the electron's transverse 
position at the time of its passing the slit plate is definite with respect to the 
alternative defined by $D$ but indefinite with respect to the alternative ``${\bf 
e}_B$ or ${\bf e}_C$''.

The above definition of positional indefiniteness makes reference to objective 
probabilities, and such probabilities, as we have just seen, can be assigned only to 
counterfactuals. The very possibility of assigning objective probabilities to the 
possible results of an unperformed measurement entails that unmeasured 
observables lack values, and that the possessed values of quantum-%
mechanical observables are extrinsic. The indefiniteness of a contingent 
property thus entails the property's extrinsic nature. The indefiniteness of the 
physical world makes it necessary to conceive of its contingent properties as 
extrinsic, or supervenient on property-indicating facts. A position can be 
indefinite only because (i)~it dangles from what happens or is the case in the 
rest of the world and (ii)~what happens or is the case in the rest of the world 
may not be enough to determine its precise value. If the positions of material 
objects were not taken from (defined by) position-indicating facts (that is, if 
they were intrinsic), they would have to be taken from an intrinsically 
differentiated space. But an intrinsically differentiated space is an infinitely 
differentiated space, and such a space has no room for indefinite values. Thus 
unless we are willing to take seriously the extrinsic nature of quantum-%
mechanical properties, we shall not be in a position to make proper sense of 
the indefiniteness that is the hallmark of quantum mechanics, and hence of 
quantum mechanics itself.

One can always conceive of a partition $\{R_i|i=1,\ldots,n\}$ into regions that 
are so small that the following is true for any specified time and any object $O$ 
other than $\cal O$: The position of $O$ relative to $\cal O$ is indefinite with 
respect to some subset $\{R_k(O)|k=1,\ldots,m\}$ of $\{R_i\}$. This is the same 
as saying that there is a subset $\{R_k\}$ of $\{R_i\}$ such that the conditionals 
$\{Q\Rightarrow O(t)\subset R_k|k=1,\ldots,m\}$ are counterfactuals and the 
probabilities associated with these counterfactuals are positive for at least two 
$k$. And this is the same as saying that no two objects ever have a definite 
relative position. A pair of material objects could have an exact relative position 
only if there existed material objects capable of indicating an exact relative 
position, but such objects do not exist. Facts never warrant the possession of a 
``sharp'' relative position. However, there are objects, which I will call 
``macroscopic'', the relative positions of which are not {\it manifestly} indefinite. 
By a {\it macroscopic object} $M$ I mean an object that 
satisfies the following criterion: Every factually warranted inference to the 
position of $M$ (relative to any other macroscopic object $\cal M$) at any 
specified time $t$ is predictable on the basis of factually warranted inferences 
to (i)~the positions of $M$ (relative to $\cal M$) at earlier times and (ii)~the 
positions of other objects (relative to $\cal M$) at $t$ or earlier times.

Let me say this again. Every position measurement that ever has been or 
will be performed on $M$ (that is, every matter of fact that has a bearing on the 
position of $M$ relative to another macroscopic object $\cal M$) has a range of 
values $\{M\subset R_i|i=1,\ldots,n\}$ between which the 
measurement can distinguish. If we take into account every position 
measurement performed on $M$ before a time $t$ and every position 
measurement performed on every other object before or at the time $t$, and if 
the result of every position measurement on $M$ made at the time $t$ is 
predictable\footnote{%
By saying that a factually warranted inference to the position of a 
macroscopic object is predictable, I do not mean that the position-indicating 
fact is predictable, but that the position indicated by the fact is predictable. Note that 
several position measurements (with different ranges of values) can be performed 
on the same object at the same time.}
on that basis via the pertinent classical laws, and if this is the case 
for every time $t$ at which a position measurement is performed on $M$, then, 
and only then, $M$ is a macroscopic object. In this case nothing ever 
indicates a departure from what is 
predictable on the basis of the pertinent classical laws and earlier
position-indicating facts. When I say that the positions of macroscopic objects 
are not manifestly indefinite, what I mean is that the indefiniteness of these 
positions is never evidenced by such a departure. Every matter of fact about 
$M$'s present position follows via the pertinent classical laws from matters of 
fact about $M$'s past positions and about the past and present positions of 
other objects.\footnote{%
Since the formal expression of the indefiniteness of an object's position refers 
to counterfactuals, evidence of positional indefiniteness cannot be direct. The 
most direct evidence we can have is the unpredictability of position-indicating 
facts. What is evidenced by the unpredictability of a position-indicating actual 
event $e$ is a {\it counterfactual} indefiniteness: the indefiniteness that would 
have obtained had $e$ not occurred, other things being equal.}

The above definition of a ``macroscopic object'' $M$ involves another 
macroscopic object $\cal M$. This is as it should be since objects are 
macroscopic by virtue of their {\it relative} positions. We may introduce a 
``macroscopic (reference) frame'' (previously called the ``laboratory frame'') 
that is riveted to any macroscopic object, and in which the position of every 
other macroscopic object is not manifestly indefinite. To see that the choice of 
reference object is immaterial, let $\{C_i\}$, $i=1,2$, be two coordinate systems 
having for their respective origins the centers of mass of two macroscopic 
objects ${\cal M}_i$. Even though the coordinate points of $\{C_2\}$ are 
somewhat fuzzy relative to $\{C_1\}$, the two sets of coordinate points are 
physically equivalent, not merely ``for all practical purposes'' but strictly, for 
there isn't any actual, physical difference matching the conceptual 
difference between them. By definition, the relative position of a pair of 
macroscopic objects is not manifestly indefinite. Hence nothing ever happens 
or is the case that would make it possible to distinguish between the two 
frames.

It is one thing to define macroscopic objects but quite another to show that 
such objects exist. There can be an unpredictable matter of fact about the position 
of $O$ at a time $t$ only if there are detectors with sensitive regions that 
are smaller than the space over which $O$'s position is distributed. (By saying 
that $O$'s position is distributed over a set $\{R_i\}$ of mutually disjoint regions, 
I mean that the prior probabilities associated with the conditionals 
$Q\Rightarrow O(t)\subset R_i$ are positive.) But detectors with sufficiently 
small sensitive regions do not always exist. There is a finite limit to the 
definiteness of the relative positions of material objects, and there is a finite 
limit to the spatial resolution of actually existing detectors. Hence there must be 
objects whose positions are the sharpest in existence. The position of such an 
object cannot be manifestly indefinite, for want of detectors capable of probing 
the space over which it is distributed.\footnote{%
Note that the definition of a ``macroscopic object'' does not stipulate that 
events indicating departures from the classically predicted positions occur with 
zero {\it probability}. An object is entitled to the label ``macroscopic'' if no such 
event {\it actually} occurs. What matters is not whether such an event {\it may} 
occur (with whatever probability) but whether it ever {\it does} occur. We cannot 
be certain that a given object qualifies as macroscopic, inasmuch as not all 
matters of fact about its whereabouts are accessible to us. But we can be 
certain that macroscopic objects exist, and that the most likely reason why $M$ 
is macroscopic is the nonexistence of detectors with sensitive regions that are 
smaller than the space over which $M$'s position is distributed. ($M$ could 
also be macroscopic for the unlikely reason that such detectors, though they 
exist, never indicate a departure from $M$'s classically predicted position.)}

While the positions of things dangle from (are supervenient on)
position-in\-di\-ca\-ting facts, the positions of macroscopic objects do so in a way 
that evinces no indefiniteness. Although no object ever follows a definite 
trajectory, the positions of macroscopic objects -- ``macroscopic positions'', for 
short -- evolve in a completely predictable fashion. Applying to macroscopic 
positions the formalism of quantum mechanics is therefore completely 
uncalled-for. It is perfectly legitimate to treat these positions as forming a self-%
contained system of {\it intrinsic} positions that dangle causally from each 
other, rather than a system of {\it extrinsic} positions that dangle ontologically 
from position-indicating facts. While even a macroscopic object has a position 
only because of the facts from which this can be inferred, the dependence of 
this position on position-indicating facts is a {\it qualitative} (ontological or 
existential) dependence, not a {\it quantitative} one. For the quantitative 
purposes of physics, it is legitimate to ignore this dependence, to 
consider macroscopic positions ``in themselves'' (out of relation to position-%
indicating facts), to treat them as facts (rather than as inferences from facts), 
and to apply to them classical causal concepts. It ought to be borne in mind, 
however, that causal concepts are emergent: Causality isn't part of the 
ontological foundation. As we saw earlier in this section, all the determining 
that goes on in the physical world is the determining {\it of probabilities} 
associated with possibilities. There aren't any causally determined {\it facts}. 
Causality, like color, lies in the mind of the
beholder~\cite{Menzies-Price1993}.\footnote{%
``The law of causality, I believe, like much that passes muster among 
philosophers, is a relic of a bygone age, surviving, like the monarchy, only 
because it is erroneously supposed to do no harm.'' -- Bertrand 
Russell~\cite{Russell1913}.}

\section{Interpreting the Copenhagen Interpretation}

Much of what has been said in the last two sections hinges on the following 
questions: What constitutes a (matter of) fact? What is an actual event or 
state of affairs? We are now in a position to answer these questions. As 
we have just seen, it is legitimate to ignore the extrinsic nature of macroscopic 
positions, to treat them as facts. The relevant facts, actual events, or actual 
states of affairs either are macroscopic positions or are definable in terms of 
such positions. It is irrelevant that the property-indicating position of a 
macroscopic pointer needle dangles ontologically from facts that involve other 
macroscopic positions. Where the indicated property is concerned, the indicating 
position can be thought of as intrinsic, and hence as a position-indicating fact.\footnote{%
An apparatus pointer is not strictly a macroscopic object according to the given 
definition. In a typical measurement, nothing allows us to predict the pointer's 
final position. However, {\it before} and {\it after} the measurement the pointer 
behaves as a macroscopic object, whose position can be considered 
independent of other position-indicating facts. Therefore the transition from the 
initial to the final pointer position can also be considered independent of other 
position-indicating facts, and thus as an actual event.}

Thus there exists a ``classical domain'' of (macroscopic) positions that are not 
manifestly indefinite, and that can be thought of as intrinsic or as being factual 
{\it per se}. This is fortunate, for otherwise quantum mechanics would be 
inconsistent, as its very formulation presupposes property- and time-indicating 
facts. Until recently the CIQM has been the only interpretation of quantum 
mechanics that acknowledges the logical dependence of quantum mechanics 
on a classical domain, and therefore, in my opinion, it has been the only 
interpretation worth considering. The CIQM has been censured for being 
``vague, obscure, and maybe even inconsistent'' by Loewer~\cite{Loewer1998}, 
but these strictures do not touch the core of the CIQM, which consists in the 
rejection of fundamental microscopic space-time realities and the substitution 
for them of property-indicating facts amenable to classical description. They 
address extraneous attempts to reintroduce unwarranted classical ideas, such 
as the notion of a quantum state that evolves in an intrinsically differentiated time.

According to Loewer, the vagueness of the conceptual muddle he passes 
off as the ``Copenhagen interpretation'' lies in the absence of (i)~a clear 
distinction between classical and quantum systems and (ii)~a clear definition of 
``measurement''. Its obscurity consists in its ``spooky'' nonlocality, in the 
``relationship between measurement and determinate reality'', and in our (or at 
least, Loewer's) inability to understand how the position of a particle can lack a 
determinate value. Its possible inconsistency is that (according to Loewer) ``it 
makes assertions about the nature of quantum-mechanical reality'' and at the 
same time ``denies that anything can be known about that reality''.

As I said, none of these strictures touch the core of the CIQM. In my 
forthcoming~\cite{Mohrhoff2000} I have expanded this core into a complete ontological 
interpretation, dubbed the ``Pondicherry interpretation of quantum mechanics'' 
(PIQM). The PIQM makes a clear distinction between the 
classical and quantum domains. The classical domain contains all those 
possessed properties that are not manifestly indefinite (including those that 
warrant inferences to possessed properties in the quantum domain). The 
quantum domain contains all other factually warranted properties. The 
term ``measurement'' is likewise clearly defined. A measurement is not 
something that causes the ``collapse'' of an evolving quantum state. Every 
property-indicating fact (that is, every event or state of 
affairs in the classical domain warranting an inference to some possessed 
property) qualifies as a ``measurement result''. The relationship between 
measurement and determinate reality is equally clear. Determinate reality is the 
totality of factually warranted properties, while measurements provide the property-%
indicating facts.

As to the alleged nonlocality of the CIQM, it depends. If this is supposed to 
mean that ``a measurement of a part of a system at one location can 
instantaneously change the physical situation of far distant parts of that 
system''~\cite{Loewer1998}, then it does not exist. This kind of action at a 
distance is one of the absurdities that follow from the spurious notion of an 
evolving quantum state. According to both the PIQM and the solid core of the 
CIQM, all that the quantum formalism tells us about is objective correlations 
between possibilities and statistical correlations between factually warranted 
properties or property-indicating facts -- diachronic correlations between the 
factually warranted properties of the same system at different times and 
synchronic correlations between the factually warranted properties of different 
systems in spacelike separation. It says nothing whatsoever about causal 
connections between the correlated properties or facts. In particular, it says 
nothing about changes in physical situations that are caused by 
measurements~\cite{Mohrhoff2000}. On the other hand, the very fact that 
quantum mechanics is inconsistent with local action, as was shown in Sec.~5, 
implies its nonlocality. Nonlocality is a characteristic of the observed 
correlations (the diachronic as well as the synchronic ones) and thus 
independent of any particular explanation of the correlations (such as an 
instantaneous action at a distance).

As to the ``spookiness'' of the nonlocality evinced by the 
correlations, it is a subject for psychology or neurophysiology rather than 
physics, and so is Loewer's inability to understand how the position of a 
particle can lack a determinate value. These nonphysical issues deserve a 
separate section (Sec.~10). That a sense of the miraculous accompanies all 
fundamental explanations is to be expected, however. Explanations begin with 
the fundamental behavior of matter, which can be described, but which cannot 
itself be explained. (If it could, it wouldn't be fundamental.) Diachronic 
correlations that are not manifestly indeterministic can be passed off as causal 
explanations. But when we deal with synchronic correlations or diachronic 
correlations that are manifestly indeterministic, causal concepts are out of 
place. Trying to causally explain these correlations is putting the cart in front of 
the horse. It is the correlations that explain why causal explanations work to the 
extent they do. They work in the classical domain where we are dealing with 
macroscopic objects, and where the correlations between property-indicating 
facts evince no statistical variations (dispersion). If we go beyond this domain, it 
becomes clear that even where no statistical variations are in evidence, the 
correlations between facts are statistical rather than causal.

Loewer's claim that the CIQM both ``makes assertions about the nature of 
quantum-mechanical reality'' and ``denies that anything can be known about 
that reality'' is based on the following claims, which he attributes to the CIQM: 
``The right way to understand quantum mechanics is not as a true description 
of physical reality but rather as an {\it instrument} for predicting the outcomes 
of laboratory experiments. There is no coherent interpretation of the
quantum-mechanical formalism as describing an unobservable reality that is 
responsible for those experimental results. That reality is forever beyond our 
ken.'' The PIQM denies not merely the possibility of such an interpretation but 
the very existence of an unobservable reality that is responsible for the 
experimental results. There is no reality beyond our ken. Quantum mechanics is an 
instrument for assigning conditional probabilities to possible property-indicating facts 
on the basis of actual property-indicating facts. Under certain conditions, specified above, 
these conditional probabilities are objective and indicative of an objective indefiniteness. 
This has implications concerning the actual spatiotemporal differentiation of the physical 
world. And all this is part of the true and complete description of physical reality that 
quantum mechanics affords.\footnote{%
Saying that quantum mechanics affords a true and complete description of physical reality 
is obviously very different from claiming that the {\it state vector} is such a description.}
If there is anything that is incomplete, it is the physical world, but its 
incompleteness exists only in relation to a conceptual framework that is more 
detailed than the physical world, as I proceed to show.

\section{The Spatiotemporal Differentiation of the Physical \\World}

Because there exists a finite limit to the spatial resolution of actually existing 
detectors, any finite region $R$ contains at most a finite number of regions 
$R_k$ the distinctness of which is physically realized. Accordingly, for any material object 
$O$ located within $R$ at most a finite number of alternative 
positions (``inside $R_k$'') are available as possible attributes. Hence there 
exists in our minds (that is, we can conceive of) a finite partition $\{R_i\}$ of the 
macroscopic frame that exists {\it only} in our minds. The 
elements of $\{R_i\}$ are so small that there aren't any detectors capable of 
realizing their conceptual distinctness. Nothing in the physical world 
corresponds to the distinctions we make between those regions.

Suppose that $\{R_i\}$ is a partition of $R$ at the limit of resolution achieved by 
actually existing detectors. How should we visualize this partition? Certainly 
not as a set of sharply bounded regions! Sharp boundaries imply exact 
positions, and exact positions imply 
the existence of detectors with infinitesimal sensitive regions, in 
contradiction to the finite actual differentiation of $R$. Since no detector ever 
has a sharply bounded sensitive region, no object ever possesses an exact 
position, and since no object ever possesses an exact position, no detector 
ever has a sharply bounded sensitive region. Even the boundaries of 
macroscopic detectors are fuzzy. But they are fuzzy only in relation to a finer 
partition of $R$, and this exists solely in our minds. Thus not only the mental 
picture of sharply bounded regions $R_i$ but also the mental picture of fuzzily 
bounded regions $R_i$ is more detailed than the finitely differentiated reality it 
is supposed to represent. The boundaries of these regions are neither sharp 
nor fuzzy! The notion that ``sharp'' and ``fuzzy'' are jointly exhaustive terms 
originates in an inadequate theoretical representation of the world's actual 
spatial differentiation. It involves a conception that is inconsistent with quantum 
mechanics -- the conception of an intrinsically and infinitely differentiated 
space.

While no object ever has a definite position, the positions of macroscopic 
objects are not manifestly indefinite. How real is the indefiniteness of a position 
that is not manifestly indefinite? The answer is, {\it not real at all}. In a world 
that is spatially differentiated only to the extent that spatial relations and 
distinctions can be inferred from facts, no object has a sharp position. But 
there are objects that have the sharpest positions in existence, and the position 
of such an object is not fuzzy in any actual sense. For its fuzziness to be 
actual in some sense, the position has to be distributed, either statistically or 
counterfactually, over the sensitive regions of actually existing detectors. But if 
it were distributed over such regions, it would not be among the sharpest 
positions in existence -- the positions of the detectors would be sharper. Thus the 
positions of macroscopic objects are fuzzy only in relation to an unrealized 
degree of spatial differentiation. The indefiniteness of a macroscopic position 
exists only in relation to a theoretical framework that is more detailed than the 
physical world.

As positions have no physical reality unless they are attributable to material 
objects, so times have no physical 
reality unless they are attributable to possessed properties, as the times at 
which these are possessed. The reason 
this is so is that time, like space, isn't something that is intrinsically 
differentiated. It takes time-indicating 
facts to differentiate the world timewise. But times are 
indicated by factually warranted relative positions. Since these are 
more or less indefinite, the indicated times, too, are more or less indefinite. For 
instance, if there isn't any matter of fact about the time $T_s$ at which an 
electron passes through the slit plate, $T_s$ lacks a definite value with respect 
to any partition of the interval between the time of emission by the electron gun 
and the time of detection behind the slit plate.\footnote{%
The use of a capital $T$ instead of either a lower-case $t$ or the operator 
symbol $\hat T$ is intended as a reminder that time is not a proper
quantum-mechanical observable (a Hermitian operator on a Hilbert space), and 
that $T_s$, consequently, is not one of the possible values of such an observable. In 
general, neither a self-adjoint time operator conjugate to the 
Hamiltonian~\cite{Pauli1958, Allcock1969} nor a time-of-arrival 
operator~\cite{Oppenheimetal1998, Oppenheimetalsub, Aharonovetal1998} 
exists. This is a consequence of the fact that quantum-mechanical probability 
distributions are distributions over the possible results of possible 
measurements that are actually or counterfactually performed at {\it specified} 
times. There are no quantum-mechanical probability distributions over time 
intervals.}
However, there are {\it macroscopic times} that are not manifestly indefinite, 
just as there are macroscopic positions the indefiniteness of which is never 
evidenced by departures from what is predictable on the basis of classical laws 
and earlier property-indicating facts.

Let us extend our definition of the ``macroscopic frame'' to include the not 
manifestly indefinite times indicated by macroscopic positions. It follows from 
what has just been said that there exists in our minds (that is, we can conceive 
of) a partition of the macroscopic frame into finite intervals of time that are so 
small that nothing in the physical world corresponds to the distinctions we 
make between these intervals. Once again the distinctions exist solely in our 
minds. The physical world is temporally differentiated only to the extent that 
temporal relations and distinctions can be inferred from facts, and the facts 
warrant neither the inference of a definite temporal relation nor the partition of 
any finite interval of physical time into infinitely many time spans. As there is a 
finite limit to the spatial resolution of actually existing detectors, so there is a 
finite limit to the temporal resolution of actually existing clocks\footnote{%
There is another reason why during a finite time span $\tau$ a material object 
can possess at most a finite number of distinct time-indicating properties. Let 
$Q_C$ be an observable whose eigenkets represent the distinct 
time-indicating properties of a clock $C$. If $Q_C$ were measured an infinite 
number of times during $\tau$, $C$ would not function as a clock, for the result 
would always be the same~\cite{Misra-Sudarshan1977, Chiu-Sudarshan1977, 
Peres1980b}.}~\cite{Peres1980a}.
No property is ever possessed at a definite time. But there are clocks, which we 
may call ``macroscopic clocks'', that indicate the sharpest times in existence, 
which we may call ``macroscopic times''. Macroscopic times are not fuzzy in 
any actual sense; they are fuzzy only in relation to an unrealized degree of 
temporal differentiation. Like the indefiniteness of a macroscopic position, the 
indefiniteness of a macroscopic time exists only in relation to a theoretical 
framework that is more detailed than the physical world.

The seemingly intractable problem of understanding quantum mechanics is a 
consequence of our dogged insistence on obtruding onto the physical world a 
theoretical framework that is more detailed than the physical world. We have 
this inveterate tendency of building reality ``from the bottom up''. Atomizing is 
the way we naturally think. As Wilson~\cite[p.~50]{Wilson1998} put it, ``[t]he 
descent to minutissima\ldots\ is a driving impulse of Western natural science. It 
is a kind of instinct.'' Not only do we atomize matter (which has its physical 
legitimacy) but we also atomize space and time (which is a mistake), and we 
tend to model the atomization of matter after the atomization of space (which is 
another mistake). That is, we tend to think of the parts of matter as being 
defined by the parts of space. In actual fact, the ultimate ``parts'' of matter are 
the fundamental particles, and these are not defined by the ``parts'' of space. 
Space isn't something that has parts, so it cannot serve to define parts. The 
``parts'' of matter are defined by the spatial relations that exist between them. 
This view, and only this, allows the spatial relations to possess indefinite 
values.

The fundamental particles exist in space only in the sense that (more or less 
indefinite) relative positions can be attributed to them. Considered in itself, out 
of relation to other material objects, a fundamental particle does not exist in 
space, for in itself it has neither a position nor a form. Physical space exists {\it 
between} the fundamental particles; they {\it unfold} it by means of their 
relative positions; it is {\it spanned} by them. Moreover, while multiplicity is 
attributable to the fundamental particles {\it qua} spatial relata, it is not 
attributable to the fundamental particles considered out of relation to each 
other. Reality, therefore, is built ``top-down'': By entering into a multitude of 
spatial relations with itself, $\cal E$ takes on not only the aspect of a spatially 
differentiated world but also the aspect of a multiplicity of fundamental 
particles. And by allowing the relations to {\it change}, to possess different 
values {\it in succession}, it takes on the further aspect of a temporally 
differentiated world.

There are limits to the resulting differentiations. While the 
sharpest relative positions and times are indefinite only relative to an unrealized 
degree of differentiation, the more fuzzy ones are indefinite relative to the realized 
degree of differentiation (that is, they are manifestly indefinite). But manifestly 
indefinite positions and times cannot be thought of as intrinsic. There is therefore 
another sense in which reality is built ``from the top down''. The 
positions of things are defined in terms of the not manifestly indefinite relative 
positions of macroscopic objects, which can be thought of as being factual by 
themselves. The times at which properties are possessed are defined in terms 
of the not manifestly indefinite times indicated by macroscopic positions, which 
also can be thought of as being factual by themselves. The positions of 
the microscopic {\it parts} and the times at which they are possessed thus dangle from 
(are supervenient on) the property- and time-indicating positions of macroscopic {\it wholes}.

\section{Stuff and Nonsense}

What is it that prevents us from coming to terms with quantum mechanics without 
extraneous additions like hidden variables~\cite{Bohm1952} or spontaneous 
collapses~\cite{Ghirardietal1986, Pearle1989, Pearle1997}, without 
using ``world'' in the plural~\cite{DeWitt-Graham1973}, and without implicitly or 
explicitly distinguishing between a mind-constructed ``internal'' or ``empirical'' reality 
and a mind-independent ``external'' or ``veiled'' reality \cite[p.~113]{Putnam1988}% 
\cite{dEspagnat1995} or dragging in consciousness or knowledge in other 
ways \cite{Lockwood1989, Albert1992, vNeumann1955, London-Bauer1983, Peierls1991, 
Page1996, Stapp1993, Mermin1998}? It 
is the {\it id\'ee fixe} that the world's synchronic multiplicity is founded on the 
introduction of surfaces that carve up space in the manner of three-%
dimensional cookie cutters. I call it the ``cookie cutter paradigm'' (CCP).

If the physical reality of space-dividing surfaces is accepted, it needs to be 
accounted for, and there are at least three ways of doing so. The first is to 
understand those surfaces as boundaries between ``full'' and ``empty'' space 
(that is, as closed surfaces encompassing some kind of stuff). This is the way 
of the Greek atomists, who taught that atoms are filled with being while the 
empty space around them lacks being. The second -- the most literal version of 
the CCP -- is due to Plato. Plato's Forms have an immaterial existence of their 
own, independently of their instantiation in the physical world. Insofar as they 
bear spatial connotations, they connote closed surfaces. Divisions in material 
space exist to the extent that Forms with spatial connotations are present 
(instantiated) in the physical world.

The third way of accounting for the existence of spatial divisions is to attribute 
them to space itself. On this account, all conceivable divisions of space are 
physically real and intrinsic to space. Reduced to one dimension this leads to 
the view that a set of points in one-to-one correspondence with the real 
numbers is intrinsic to a continuous line. This raises the issue of whether these 
points ``make up'' or ``fill'' the line or are separated by infinitesimal intervals. It 
is customary to equate real numbers whose decimal expansions converge, 
e.g., $0.4\bar9=0.5$. If we follow this practice -- not everyone 
does~\cite[p.~263]{Rucker1983}~-- then there aren't any infinitesimal ``gaps'' 
between the points on a line.\footnote{%
Consider a line segment $L$ with boundary points labeled $0$ and $1$, 
respectively. Next consider the numbers in the interval $I=(0,1)$ that have a 
binary expansion of up to $n$ digits. These numbers have the general form 
$0.[n]$, where $[n]$ is a string of $n$ digits ($0$'s or $1$'s). The points 
corresponding to these numbers divide $L$ into $2^n$ segments of length 
$l_n=2^{-n}$. In the limit $n\rightarrow\infty$ we obtain all real numbers in $I$. 
Let $s$ be any real number and let $(s_n)$ be a sequence of real numbers 
such that $s-s_n=2^{-n}$. If $\lim_{n\to\infty}s_n=s$, so that in particular (in 
binary notation) $0.0\bar1=0.1$, the assumption that there are infinitesimal line 
segments between the points on $L$ corresponding to the real numbers in $I$ 
leads to a contradiction, for the upper and lower boundaries of those segments 
are defined by the same number.}
But if we interpret real numbers as points on a continuous line, or use them to 
label such points, then it is legitimate to consult the visual image of a 
continuous line, and to demand that the properties we attribute to the real 
numbers be consistent with it. And arguably the practice of equating 
numbers with convergent decimal expansions is not consistent with the 
continuity of a line in phenomenal space.\footnote{%
Let $\{n\}$ denote a string of $n$ $1$'s. (We again use binary notation.) If 
we visualize the line segment $L_n$ corresponding to the interval $(0.0\{n\},0.1)$, and if we visualize the right end of $L_n$ enlarged by a factor $2^m$ 
with every increase of $n$ by $m$, then the segment $L_{n+m}$ 
corresponding $(0.0\{n+m\},0.1)$ looks exactly the same as $L_n$. And 
this ought to be equally true of the segment $L_\infty$ corresponding to 
$(0.0\bar1,0.1)$.}
Be that as it may. What is clear is that the idea that all conceivable divisions of 
phenomenal space are intrinsic to physical space, leads naturally to the notion 
that physical space either contains or is identical with the set \R.

Physicists have long since discarded the Democritean notion that the basic 
material constituents are closed surfaces filled with continuous being. If they 
ever entertained the Platonic notion that space-dividing surfaces owe their 
physical reality to Forms that have a reality {\it ante rem}, they have long since 
discarded this notion as well. What remains to be discarded is (i)~the notion 
that space-dividing surfaces are physically real, and (ii)~the conception of 
space to which this notion leads if one rejects the Democritean and Platonic 
accounts. Sharp space-dividing surfaces exist solely in our minds. They do not 
exist as the forms of material objects, for the forms of material objects are sets 
of more or less indefinite relative positions between formless entities, rather 
than bounding surfaces (Sec.~3). Nor are space-dividing surfaces 
intrinsic to physical space, as 
the behavior of electrons in two-slit interference experiments amply 
demonstrates (Sec.~5).

To bring home just how insidiously the CCP prevents us from making sense 
of quantum mechanics, let us examine some of its implications. To begin with, 
the idea that synchronic multiplicity depends on a partition of space into 
mutually disjoint regions implies the prior existence of a spatial expanse that 
gets partitioned or contains the partitions. The CCP thus prejudices us in favor 
of {\it substantivalism}, a doctrine that we found to be inconsistent 
with quantum mechanics. In conjunction with the CCP, substantivalism implies 
that not only space but {\it each part of space} is a separate constituent of the 
world. If the parts of matter exist by virtue of the parts of space, spatial 
divisions cannot arise from (processes involving) matter; they must be inherent in 
a preexistent space.

What transpires next depends on whether the world is or is not infinitely 
divided spacewise. If the division of space ends with the creation of finite 
bounded regions, as in the respective theories of Democritus and Plato, it 
seems inevitable that we follow these philosophers in attributing the existence 
of bounded regions to the existence of material objects, and thus conceive of 
existing boundaries as forms of material objects. But then the following 
question arises: Why can't it happen that different material objects overlap? 
The answer to this question is, because the CCP defines synchronic 
multiplicity in terms of geometrical divisions. Suppose that there exist two 
bounded regions $A$ and $B$ having a finite intersection $C=A\cap B$. Then it 
is not the case that there exist two material objects whose respective forms are 
the boundaries of $A$ and $B$. Instead there exist three such objects whose 
respective forms are the boundaries of $A-B$, $B-A$, and $C$. The object 
occupying $C$ is {\it one} object (which may be a part of the object occupying 
$A$, or a part of the object occupying $B$, or a part of the object occupying 
$A\cup B$), rather than {\it two} objects (a part of the object occupying $A$ {\it 
and} a part of the object occupying $B$). Thus it is logically impossible for two 
objects to overlap (that is, for a part of one object to occupy the same region of 
space as a part of another object).

Material objects, however, move. This is something that bounded regions of 
space {\it per se}, considered out of relation to time, cannot do, and this raises 
a further question. We know that two material objects cannot ``overlap''. If it 
seemed as if they did, their apparent intersection would contain a part of either 
object rather than a part of each object. If two identical objects came to occupy 
the same space, they would cease to be two objects. But this does not explain 
why we never see a part of one object become numerically identical with a part 
of another object, or two identical objects merge into one object. The obvious 
``explanation'' of this is that material objects are not only bounded by surfaces 
but also ``filled to capacity'' with some continuous stuff.

Physics offers a different account of the apparent impenetrability of material 
objects: a repulsive force. The physical reason why two material objects $M$ 
and $N$ lacking common parts cannot come to occupy the same space is that 
their respective parts repel each other. This explanation involves the spatial 
relations, or the distances, between the parts of $M$ and the parts of $N$, as 
well as a force opposing attempts to reduce those distances. It further involves 
the parts of $M$ and the parts of $N$, but only as the relata of those spatial 
relations. It does not involve them as bounded regions filled with stuff. It 
involves neither the forms of the parts nor any spatially extended stuff. Where 
physics is concerned, space-filling stuff and forms {\it qua} bounding surfaces 
are explanatorily irrelevant. They are artefacts of the CCP and the assumption 
that the division of space ends with finite bounded regions. In reality, as 
described by quantum mechanics, there are no bounding surfaces. Forms are 
made of spatial relations between formless parts. And if there is stuff, it consists 
in nothing but the spatial relata the existence of which is implied by the spatial relations.

If the CCP is combined instead with the assumption that the division of space 
never ends, or ends with infinitesimal regions or with a set of points cardinally 
equal to the reals, it leads us up a different garden path. In this case all points 
or infinitesimal regions are separate constituents of the world, and all physical 
properties are locally instantiated -- they are properties of those points or 
infinitesimal regions. The form of an ordinary material object then consists of 
spatial relations between locally instantiated physical properties. If we take into 
account that a generic 
material object is composed of a finite number of noncomposite entities, we are 
led to conclude that the form of such an object is made up of the spatial 
relations between physical properties that are instantiated at a finite number of 
points or infinitesimal regions. It stands to reason that these locally instantiated 
physical properties are the characteristics of a particle species: mass, spin, and 
charges. Note that a pointlike form is not contained in this list of properties. 
Saying that a fundamental particle is pointlike is the same as saying that its 
properties (not including a form) are instantiated at a point or an infinitesimal 
region of space.

It is clear that the spatial relations between these locally instantiated properties cannot be 
indefinite. The CCP thus makes it impossible to understand how the position of 
a particle can lack a determinate value. If the synchronic multiplicity of the world conformed 
to the CCP, space would be intrinsically and infinitely divided, a noncomposite 
object could not but exist at a definite point of space or inside a definite 
infinitesimal region, and the distances between such objects would necessarily 
be sharp.\footnote{%
Recall that the coordinates presupposed by quantum mechanics have a direct 
metric significance. The ``points of space'' being Cartesian coordinate points, 
their distances are determinately related to the differences between their 
coordinates.}
If, as Albert~\cite[p.~11]{Albert1992} has claimed, the behavior of electrons in 
two-slit experiments is ``quite unlike what we know how to think about'', it is 
because we labor under the delusion of the CCP. If ``[n]obody knows how it 
can be like that'',\footnote{%
``I think it is safe to say that no one understands quantum mechanics\ldots. Do 
not keep saying to yourself, if you can possibly avoid it, `But how can it be like 
that?' because you will go `down the drain' into a blind alley from which 
nobody has yet escaped. Nobody knows how it can be like that'' -- 
R.P. Feynman~\cite[p.~129]{Feynman1967}.}
it is because everybody is deluded by the CCP. If we could accept that the 
synchronic multiplicity of the world is based instead on spatial relations, we 
would have no reason to suppose that spatial relations must have determinate 
values. On the contrary, taking into account that a finite number of constituents 
is sufficient for the existence of any known material object, we have reason to 
suppose that the relative positions of these constituents {\it cannot} have 
determinate values. In fact, we {\it know} that what ``fluffs out'' matter is not a 
repulsive force but the indefiniteness of the relative positions of its constituents 
(in conjunction with the Pauli exclusion principle and the fact that those 
constituents are fermions). If it were not for this indefiniteness -- other things 
being equal -- none of the familiar objects around us would exist.

The CCP can lead to worse. This happens when the quantum state -- by 
definition a system of probability distributions -- is construed not only as an 
instantaneous state of affairs $\psi(t)$ that evolves in an infinitely differentiated 
time but also as a local state of affairs $\psi({\bf r})$ that assigns physical 
properties to the point~$\bf r$. This is done, for instance, by 
Albert~\cite[p.~126]{Albert1992}, who sets out with the following bundle of assumptions: 
``Suppose that there's just one world. And suppose that there's just one 
complete story of the world that's true. And suppose that quantum-mechanical 
state vectors are complete descriptions of physical systems. And suppose that 
the dynamical equations of motion are always [and, presumably, everywhere] 
exactly right.'' The first assumption excludes the many-worlds 
interpretation~\cite{DeWitt-Graham1973}; the second rules out the consistent-%
histories philosophy~\cite{Griffiths1984,GellMann-Hartle1990, Omnes1992}; the 
third eliminates Bohmian mechanics~\cite{Bohm1952} and spontaneous 
collapse theories~\cite{Ghirardietal1986, Pearle1989, Pearle1997}. The four 
assumptions together then lead to the many-minds
interpretation~\cite{Albert1992, Albert-Loewer1988}, or so it is suggested by 
its proponents. Common to all of these attempts to make sense of quantum mechanics 
is the idea that the dynamical equations -- in the simplest case, the 
Schr\"odinger equation; in the case of spontaneous collapse theories, a 
modified, stochastic equation -- are {\it always and everywhere} exactly obeyed, 
where ``always'' and ``everywhere'' stand for ``at every instant of time'' and ``at 
every point of space'', respectively. Since the physical world is not infinitely 
differentiated spacewise or timewise, these usual acceptations of ``always'' and 
``everywhere'' have no meaningful application to the physical world. For this 
reason alone all of the above attempts to make sense of quantum mechanics 
are fatally flawed.

The CCP has neurophysiological underpinnings. We are adept at recognizing 
three-dimensional objects in drawings that contain only outlines. (In fact, we 
can't help but perceive three-dimensional objects. We always see a Necker 
cube as pointing either in or out.) Why is it so easy to recognize an outline as 
an object? The answer is, because of the way the brain processes visual 
information. The seminal work of Hubel and Wiesel~\cite{Hubel-Wiesel1979} 
supports the following account. Visual information flows from retinal receptor 
cells via retinal ganglion cells to either of two lateral geniculate nuclei, and on 
to the primary visual cortex. The receptive field of each retinal ganglion or 
geniculate cell is divided into either an excitatory center and a concentric 
inhibitory surround (the ``on center'' configuration) or the reverse configuration 
(``off center''). (The group of retinal receptor cells from which a retinal ganglion 
or geniculate cell receives input is known as the cell's receptive field.) Thus an 
``on center'' cell responds best to a circular spot of light of a specific size, 
responds well to a bright line that just covers the center (since then most of the 
surround is not covered by the line), and does not respond at all if both center 
and surround are fully and equally illuminated.

When visual information reaches the visual cortex, two major transformations 
take place. One leads to the fusion of input from both eyes, the other to a 
rearrangement of incoming information so that most of its cells respond to 
specifically oriented line segments. The optimal stimulus may be a bright line 
on a dark background or the reverse, or it may be a boundary between light 
and dark regions. One group of orientation-specific neurons responds best to 
lines with just the right tilt in a particular part of the visual field. Another group 
of neurons, receiving input from the first group, is less particular about the 
position of the line and responds best if the line is swept in a particular 
direction across the visual field.

These data indicate that the visual representation of a physical environment 
arises by way of an analysis of the visual field that is based on contrast 
information from boundaries between homogeneously lit regions. No sense 
data arrive from regions that are homogeneously colored and evenly lit. The 
interior of such a region is {\it filled in} on the basis of contrast information 
stemming from its {\it boundary}. This explains why outline drawings are readily 
recognized as objects: The brain adds surfaces to outlines in the same way as 
it adds (unperceived) colored surfaces to (perceived) changes in color and 
brightness across edges. It also explains why the blind spot is not perceived if 
it falls inside a homogeneous region (no sense data arrive from such a region 
anyway), and why color perception is so remarkably faithful to the reflectances 
of colored surfaces, and correspondingly insensitive to the spectral 
composition of the actual radiances of such 
surfaces~\cite{Land1977}.\footnote{%
If color perception is based on discontinuous color changes across edges, 
continuous variations in illumination across the visual field go unperceived.}
And most importantly, the manner in which visual information is analyzed by 
the visual cortex explains why the synchronic multiplicity of the {\it 
phenomenal} world conforms to the CCP: Unlike the physical world, the 
phenomenal world {\it is} constructed from boundaries, and these are filled in 
with qualia. A trivial consequence of this is that two objects in the phenomenal 
world, unlike two objects in the physical world, cannot be at the same place: 
The existence of two objects in phenomenal space implies the existence of at 
least one separating boundary.

The phenomenal world differs from the physical world not only in that it is 
spatially differentiated in conformity with the CCP, but also in that it is 
intrinsically differentiated. The visual field is differentiated not only extrinsically 
by differences in perceived content but also intrinsically by the retinal receptor 
cells -- it is inherently grainy. That we are unaware of this graininess is another 
consequence of how the brain processes visual information. Recall that no 
sense data arrive from uniformly colored regions of the visual field. Such 
regions are filled in, and are filled in smoothly or homogeneously.\footnote{%
Since this involves the transition from objective brain mechanisms to subjective 
visual percepts, just how the filling in is accomplished is presently as 
impenetrable as the question of how anything material can be conscious in the 
first place.}
The inherent graininess of boundaries is similarly glossed over. Cells that 
respond to specifically oriented line segments in a particular part of the visual 
field receive input from cells that have circular receptive fields with centers 
lying along a straight line. The information coming from the latter type of cell is 
grainy; the line segment that is perceived when the former type of cell is 
stimulated, is not. The phenomenal world thus is intrinsically a world of sharp 
and continuous boundaries filled with homogeneous content, while the 
physical world is intrinsically a world of fuzzy spatial relations between relata 
that, but for their relations, are numerically identical. There could hardly be a 
greater difference between the two.

Many thinkers have been intrigued, and justifiably so, by the ability of the 
human mind to reproduce the physical world as faithfully as it does, or seems 
to do, using nothing but logic and mathematics.\footnote{%
``Difficult though it be to imagine physics either completable or incompletable, 
it is perhaps even more difficult to imagine that physics should be possible at 
all'' --~von Weizs\"acker~\cite[p.~174]{vWeizsacker1980}. ``The most 
incomprehensible thing about the world is that it is comprehensible'' --%
~Einstein~\cite[p.~112]{Schilpp1949}.}
The success of physics is indeed astonishing, but the difficulties we face in 
understanding quantum mechanics reveal that we are not all that well-%
equipped mentally. As McGinn~\cite{McGinn1995} has stressed, ``[w]e are, 
cognitively speaking as well as physically, spatial beings {\it par excellence}: 
our entire conceptual scheme is shot through with spatial notions, these 
providing the skeleton of our thought in general.'' The trouble is that our 
neurophysiological make-up conditions these spatial notions to conform to the 
CCP. Recall from note~1 that the neural processes which produce 
visual percepts and those which produce visual images are to some extent the 
same. As a consequence, every thought about the physical world that involves 
visual imagery is invariably deceived by some of the neural processes on which 
it depends.

To conclude, in order to make sense of quantum mechanics, we must detach 
our spatial notions from visual imagery. We must learn to conceive of formless 
entities. We must disregard the intrinsic multiplicity of phenomenal space. We 
must think of spatial relations as ontologically prior both to forms and to the 
multiplicity of the corresponding relata. (Without spatial relations, all there is is 
existence itself. This takes on the appearance of a world of forms when it 
enters into spatial relations with itself.) And we must counter the inherent 
definiteness of visual percepts and images by resorting to probabilistic 
concepts and contrary-to-fact conditionals. Last but not least, we must desist 
from conceiving individuation along lines laid down by the CCP.

According to Strawson~\cite{Strawson1959}, the logical distinction between 
particular and universal, and hence between subject and predicate, is founded 
on spatial distinctness. ``We regard $x$ and $y$ as distinct particular instances 
of the same universal $P$ just in so far as we acknowledge that $x$ and $y$ 
are {\it at distinct places}''~\cite[original emphasis]{McGinn1995}. Quantum 
mechanics tells us otherwise. Being at distinct places is indeed sufficient for 
being distinct instances of existence itself, but it is not necessary. What is 
necessary for being distinct instances is the existence of a spatial relation. This 
can be such that, relative to the laboratory frame, the two instances are not at 
distinct places. Nor can existence itself be thought of as a universal (Sec.~4). 
$\cal E$ is not the most general predicate but the ultimate subject. Material 
objects owe their existence to the existence of spatial relations, and these owe 
their existence not to a predicable universal but to the one existence that they 
manifoldly relate.


\begin{thebibliography}{99}
\bibitem{Pais1982}
Abraham Pais, {\it `Subtle is the Lord...': The Science and the Life of Albert 
Einstein} (Oxford: Clarendon Press, 1982).

\bibitem{Finke1980}
R.A. Finke, ``The functional equivalence in imagery and perception'', {\em 
Psychol. Rev.} {\bf 87}, No. 2, 113--132, 1980.

\bibitem{Finke-Shepard1986}
R.A. Finke and R.N. Shepard, ``Visual functions and mental imagery'', in {\em 
Handbook of Perception and Human Performance II} (New York: John Wiley \& 
Sons, 1986), edited by K.R. Boff, L. Kauffmann, and J.P. Thomas, pp. 37--1 to 
37--55.

\bibitem{Shepard-Cooper1982}
R.N. Shepard and L.A. Cooper, {\em Mental Images And Their Transformations} 
(Cambridge, MA: MIT Press, 1982).

\bibitem{Cantor1932}
Georg Cantor, in {\em Gesammelte Abhandlungen} (Berlin: Springer, 1932), 
edited by Abraham Fraenkel and Ernst Zermelo.

\bibitem{Land1977}
Edwin H. Land, ``The Retinex theory of color vision'', {\em Sci. Am.} {\bf 237}, 
No. 6, 108--128, December 1977.

\bibitem{Weyl1970}
Herrman Weyl, {\em Raum Zeit Materie}, 6th edition (Berlin: Springer, 1970); 
translation: H.L. Brose, {\it Space-Time-Matter} (London: Methuen, 1922).

\bibitem{Jackson1986}
Frank Jackson, ``What Mary didn't know'', {\em J. Phil.} {\bf 83}, 291--295, 1986.

\bibitem{Grunbaum1973}
A. Gr\"unbaum, {\it Philosophical Problems of Space and Time} (Dordrecht: 
Reidel, 1973).

\bibitem{Poincare1952}
Henry Poincar\'e, {\em Science and Hypothesis} (New York: Dover, 1952).

\bibitem{Anandan1980}
J. Anandan, ``On the hypotheses underlying physical geometry'', {\em Found. 
Phys.} {\bf 10}, 601--629, 1980.

\bibitem{Silva1997}
P.R. Silva, ``A new interpretation of the de Broglie frequency?'', {\em Phys. 
Essays} {\bf 10}, 628--632, 1997.

\bibitem{Kant1781}
Immanuel Kant, {\it Critique of Pure Reason}, first (German) edition, 1781.

\bibitem{Klauder1997}
John R. Klauder, ``Understanding quantization'', {\em Found. Phys.} {\bf 27}, 
1467--1483, 1997.

\bibitem{Klauder1998}
John R. Klauder, ``Is quantization geometry?'', {\em Comm. Math. Theo. 
Phys.} {\bf 1}, 50--64, 1998.

\bibitem{Klauder1999}
John R. Klauder, ``Metrical quantization'', in {\it Quantum Future} (Berlin: 
Springer, 1999), edited by P. Blanchard and A. Jadcyk , pp. 129--138.

\bibitem{Dirac1947}
P.A.M. Dirac, {\em The Principles of Quantum Mechanics} (Oxford: Oxford 
University Press, 1947).

\bibitem{Audi1995}
Robert Audi, {\it The Cambridge Dictionary of Philosophy} (Cambridge: 
Cambridge University Press, 1995).

\bibitem{Misneretal1973}
C.W. Misner, K.S. Thorne, and J.A. Wheeler, {\it Gravitation} (San Francisco: 
W.H. Freeman and Company, 1973).

\bibitem{Feynmanetal1965}
Richard P. Feynman, Robert B. Leighton, and Matthew Sands, {\em The 
Feynman Lectures in Physics}, Vol. 3 (Reading, MA: Addison-Wesley, 1965).

\bibitem{Lockwood1989}
Michael Lockwood, {\it Mind, Brain and the Quantum} (Oxford: Basil Blackwell, 
1989).

\bibitem{Bohm1952}
David Bohm, ``A suggested interpretation of quantum theory in terms of hidden 
variables'', {\em Phys. Rev.} {\bf 85}, 166--193, 1952.

\bibitem{Albert1992}
David Z. Albert, {\it Quantum Mechanics and Experience} (Cambridge, MA: 
Harvard University Press, 1992).

\bibitem{Lewis1986}
David K. Lewis, {\it Philosophical Papers}, Vol. II (New York: Oxford University 
Press, 1986).

\bibitem{Einstein1948}
Albert Einstein, ``Quantum mechanics and reality'', {\em Dialectica} {\bf 2}, 320-%
-324, 1948; translation in: D. Howard, ``Holism, separability and the 
metaphysical implications of the Bell experiments'', in {\em Philosophical 
Consequences of Quantum Theory: Reflections on Bell's theorem} (Notre 
Dame, Indiana: University of Notre Dame Press, 1989), edited by J. Cushin and 
E. McMullin, pp. 224--253.

\bibitem{Cassinello-SG1996}
A. Cassinello and J.L. S\'anchez-G\'omez, ``On the probabilistic postulate of 
quantum mechanics'', {\it Found. Phys.} {\bf 26}, 1357--1374, 1996.

\bibitem{Jauch1968}
J.M. Jauch, {\it Foundations of Quantum Mechanics} (Reading, MA: Addison-%
Wesley, 1968).

\bibitem{Gleason1957}
A.M. Gleason, ``Measures on the closed subspaces of a Hilbert space'', {\em J. 
of Rat. Mech. and Analysis} {\bf 6}, 885--894, 1957.

\bibitem{vKampen1988}
N.G. van Kampen, ``Ten theorems about quantum-mechanical 
measurements'', {\em Physica} {\bf A153}, 97--113, 1988.

\bibitem{Bohr1934}
Niels Bohr, {\it Atomic Theory and the Description of Nature} (Cambridge: 
Cambridge University Press, 1934).

\bibitem{Bohr1958}
Niels Bohr, {\it Atomic Physics and Human Knowledge} (New York: Wiley, 
1958).

\bibitem{Redhead1987}
Michael Redhead, {\it Incompleteness, Nonlocality and Realism} (Oxford: 
Clarendon, 1987).

\bibitem{Stapp1972}
Henry Pierce Stapp, ``The Copenhagen interpretation'', {\em Am. J. Phys.} {\bf 
40}, 1098--1116, 1972.

\bibitem{Mermin1996}
N. David Mermin, ``What's wrong with this sustaining myth?'', {\em Phys. 
Today} {\bf 49}, 11--13, March 1996.

\bibitem{Loewer1998}
Barry Loewer, ``Copenhagen versus Bohmian interpretations of quantum 
theory'', {\em Brit. J. Phil. Sci.} {\bf 49}, 317--328, 1998.

\bibitem{Peres1984}
Asher Peres, ``What is a state vector?'', {\em Am. J. Phys.} {\bf 52}, 644--650, 
1984.

\bibitem{ABL1964}
Yakir Aharonov, Peter G. Bergmann, and Joel L. Lebowitz, ``Time symmetry in 
the quantum process of measurement'', {\em Phys. Rev.} {\bf 134B}, 1410--%
1416, 1964; reprinted in {\it Quantum Theory and Measurement} (Princeton, 
NJ: Princeton University Press, 1983), edited by John Archibald Wheeler and 
Wojciech Hubert Zurek, pp. 680--686.

\bibitem{Aharonov-Vaidman1991}
Yakir Aharonov and Lev Vaidman, ``Complete description of a quantum system 
at a given time'', {\em J. Phys.} {\bf A24}, 2315--2328, 1991.

\bibitem{Reznik-Aharonov1995}
B. Reznik and Y. Aharonov, ``Time symmetric formulation of quantum 
mechanics'', {\em Phys. Rev.} {\bf A52}, 2538--2550, 1995.

\bibitem{Vaidman1998}
Lev Vaidman, ``Time-symmetrized quantum theory'', {\em Fortschr. Phys.} {\bf 
46}, 729--739, 1998.

\bibitem{Mohrhoff2000}
Ulrich Mohrhoff, ``What quantum mechanics is trying to tell us'', forthcoming in 
{\em Am. J. Phys.}; available online as ``The Pondicherry interpretation of 
quantum mechanics'', Eprint quant-ph/9903051.

\bibitem{Kastner1999a}
R.E. Kastner, ``Time-symmetrized quantum theory, counterfactuals, and 
`advanced action'$\,$'', {\em Stud. Hist. Phil. Mod. Phys.} {\bf 30}, 237--259, 
1999.

\bibitem{Kastner1999b}
R.E. Kastner, ``The three-box `paradox' and other reasons to reject the 
counterfactual usage of the ABL rule'', {\em Found. Phys.} {\bf 29}, 851--863, 
1999.

\bibitem{Vaidman1999a}
Lev Vaidman, ``Defending time-symmetrised quantum counterfactuals'', {\em 
Stud. Hist. Phil. Mod. Phys.} {\bf 30}, 373--397, 1999.

\bibitem{Vaidman1999b}
Lev Vaidman, ``Time-symmetrized counterfactuals in quantum theory'', {\em 
Found. Phys.} {\bf 29}, 755--765, 1999.

\bibitem{Vaidman1999c}
Lev Vaidman, ``The meaning of elements of reality and quantum 
counterfactuals: Reply to Kastner'', {\em Found. Phys.} {\bf 29}, 865--876, 1999.

\bibitem{Kastner2000}
R.E. Kastner, `` Comment on Mohrhoff's `What quantum mechanics is trying to 
tell us'$\,$'', Eprint quant-ph/0003098.

\bibitem{Cohen1995}
O. Cohen, ``Pre- and postselected quantum systems, counterfactual 
measurements, and consistent histories'', {\em Phys. Rev.} {\bf A51}, 4373--%
4380, 1995.

\bibitem{Menzies-Price1993}
Peter Menzies and Huw Price, ``Causation as a secondary quality'', {\em Brit. J. 
Phil. Sci.} {\bf 44}, 187--203, 1993.

\bibitem{Russell1913}
Bertrand Russell, ``On the notion of cause'', {\em Proc. Aristotelean 
Soc.} {\bf 13}, 1--26, 1913.

\bibitem{Pauli1958}
Wolfgang Pauli, {\it Encyclopaedia of Physics}, Vol. 5/1 (New York: Springer, 
1958), edited by S. Flugge, p. 60.

\bibitem{Allcock1969}
G.R. Allcock, ``The time of arrival in quantum mechanics: I.~Formal 
considerations'', {\em Ann. Phys. (NY)} {\bf 53}, 253--285, 1969.

\bibitem{Oppenheimetal1998}
J. Oppenheim, B. Reznik, and W.G. Unruh, ``Time as an Observable'', in {\it 
Proc. 10th Max Born Symposium, Wroclaw} (Berlin: Springer, 1998), 
edited by Ph. Blanchard and A. Jadczyk, pp. 204--219.

\bibitem{Oppenheimetalsub}
J. Oppenheim, B. Reznik, and W.G. Unruh, ``Minimum inaccuracy for 
traversal-time'', submitted to {\em Phys. Rev. A}, Eprint quant-ph/9801034.

\bibitem{Aharonovetal1998}
Y. Aharonov, J. Oppenheim, S. Popescu, B. Reznik, and W.G. 
Unruh, ``Measurement of Time-of-Arrival in Quantum Mechanics'', 
{\em Phys. Rev.} {\bf A57}, 4130--4139, 1998.

\bibitem{Misra-Sudarshan1977}
B. Misra and E.C.G. Sudarshan, ``The Zeno's paradox in quantum theory'', 
{\em J. Math. Phys.} {\bf 18}, 756--763, 1977.

\bibitem{Chiu-Sudarshan1977}
C.B. Chiu and E.C.G. Sudarshan, ``Time evolution of unstable states and a 
resolution of Zeno's paradox'', {\em Phys. Rev.} {\bf D16}, 520--529, 1977.

\bibitem{Peres1980b}
Asher Peres, ``Zeno paradox in quantum theory'', {\em Am. J. Phys.} {\bf 48}, 
931--932, 1980.

\bibitem{Peres1980a}
Asher Peres, ``Measurement of time by quantum clocks'', {\em Am. J. Phys.} 
{\bf 48}, 552--557, 1980.

\bibitem{Wilson1998}
Edward O. Wilson, {\em Consilience} (New York: Alfred A. Knopf, 1998).

\bibitem{Ghirardietal1986}
G.C. Ghirardi, A. Rimini, and T. Weber, ``Unified dynamics for microscopic and 
macroscopic systems'', {\em Phys. Rev.} {\bf D34}, 470--491, 1986.

\bibitem{Pearle1989}
Philip Pearle, ``Combining stochastic dynamical state-vector reduction with 
spontaneous localization'', {\em Phys. Rev.} {\bf A39}, 2277--2289, 1989.

\bibitem{Pearle1997}
Philip Pearle, ``True collapse and false collapse,'' in {\it Quantum Classical 
Correspondence} (Cambridge, MA: International Press, 1997), edited by Da 
Hsuan Feng and Bei Lok Hu, pp. 51--68.

\bibitem{DeWitt-Graham1973}
Bryce S. DeWitt and Neill Graham, eds., {\it The Many-Worlds Interpretation of 
Quantum Mechanics} (Princeton, NJ: Princeton University Press, 1973).

\bibitem{Putnam1988}
Hilary Putnam, {\it Representation and Reality} (Cambridge, MA: MIT Press, 
1988).

\bibitem{dEspagnat1995}
Bernard d'Espagnat, {\it Veiled Reality} (Reading, MA: Addison-Wesley, 1995).

\bibitem{vNeumann1955}
John von Neumann, {\it Mathematical Foundations of Quantum Mechanics} (Princeton, NJ: 
Princeton University Press, 1955).

\bibitem{London-Bauer1983}
Fritz London and Edmond Bauer, ``The theory of observation in quantum mechanics,'' in 
{\it Quantum Theory and Measurement} (Princeton, NJ: Princeton University Press, 1983), 
edited by John Archibald Wheeler and Wojciech Hubert Zurek, pp. 217--259.

\bibitem{Peierls1991}
Rudolf Peierls, ``In defence of `measurement'$\,$'', {\em Physics World} {\bf 4}, 19--20, 
January 1991.

\bibitem{Page1996}
Don N. Page, ``Sensible quantum mechanics: Are probabilities only in the mind?'', {\it Int. J. Mod. Phys.} {\bf D5}, 583--596, 1996.

\bibitem{Stapp1993}
Henry Pierce Stapp, {\it Mind, Matter, and Quantum Mechanics} (Berlin: Springer, 1993).

\bibitem{Mermin1998}
N. David Mermin, ``What is quantum mechanics trying to tell us?'', {\it Am. J. Phys.} {\bf 66}, 753--767, 1998.

\bibitem{Rucker1983}
Rudy Rucker, {\em Infinity and the Mind: The Science and 
Philosophy of the Infinite} (New York: Bantam Books, 1983).

\bibitem{Feynman1967}
Richard P. Feynman, {\it The Character of Physical Law} (Cambridge, MA: MIT 
Press, 1967).

\bibitem{Griffiths1984}
Robert B. Griffiths, ``Consistent histories and the interpretation of quantum 
mechanics'', {\em J. Stat. Phys.} {\bf 36}, 219--272, 1984.

\bibitem{GellMann-Hartle1990}
M. Gell-Mann and J.B. Hartle, ``Quantum mechanics in the light of quantum 
cosmology,'' in {\it Complexity, Entropy, and the Physics of Information} ( 
Reading, MA: Addison-Wesley, 1990), edited by W.H. Zurek, pp. 425--458.

\bibitem{Omnes1992}
Roland Omn\`es, ``Consistent interpretations of quantum 
mechanics'', {\em Rev. Mod. Phys.} {\bf 64}, 339--382, 1992.

\bibitem{Albert-Loewer1988}
D. Albert and B. Loewer, ``Interpreting the Many Worlds Interpretation'', {\it 
Synthese} {\bf 77}, 195--213, 1988.

\bibitem{Hubel-Wiesel1979}
D.H. Hubel and T.N. Wiesel, ``Brain mechanisms of vision'', {\em Sci. Am.} {\bf 
241}, 150--162, September 1979.

\bibitem{vWeizsacker1980}
C.F. von Weizs\"acker, {\em The Unity of Nature} (New York: Farrar, Straus 
and Giroux, 1980).

\bibitem{Schilpp1949}
P.A. Schilpp, ed., {\em Albert Einstein: Philosopher-Scientist} (Evanston, IL: 
Library of Living Philosophers, 1949).

\bibitem{McGinn1995}
Colin McGinn, ``Consciousness and space'', {\em J. Consc. Stud.} {\bf 2}, 220--%
230, 1995.

\bibitem{Strawson1959}
P.F. Strawson, {\em Individuals} (London: Methuen, 1959).
\end{thebibliography}
\end{document}